\documentclass[aps, prfluids, nofootinbib]{revtex4-2}
\usepackage{amsmath, amssymb, mathtools, physics, bm}
\usepackage{tikz, pgfplots}
\usepackage[svgnames]{xcolor}

\usetikzlibrary{arrows.meta, decorations.markings}

\newcommand{\bcdot}{\boldsymbol{\cdot}}
\newcommand{\bnabla}{\boldsymbol{\nabla}}

\newcommand*\circled[1]{\tikz[baseline=(char.base)]{
            \node[shape=circle,draw,inner sep=1.5pt, draw=Crimson, fill=Crimson!10, text=Crimson] (char) {$\bm{\mathsf{#1}}$};}}
\newcommand*\smlcircled[1]{\tikz[baseline=(char.base)]{
            \node[shape=circle,draw,inner sep=0.75pt, draw=Crimson, fill=Crimson!10, text=Crimson] (char) {\scriptsize $\bm{\mathsf{#1}}$};}}

\begin{document}
    \title{Surfing on chemical waves: a simple yet dynamically rich two-sphere responsive gel swimmer}

    \author{Joseph J. Webber}
    \email{joe.webber@warwick.ac.uk}
    \author{Thomas D. Montenegro-Johnson}%
    \email{tom.montenegro-johnson@warwick.ac.uk}
    \affiliation{Mathematics Institute, University of Warwick, Coventry CV4 7AL, UK}

    \date{\today}

    \begin{abstract}
        Self-oscillating gels are chemically-responsive hydrogels coupled to an oscillating chemical reaction of a stimulus solute. In response to the oscillating solute concentration field, responsive gels periodically swell and deswell, expelling their adsorbed water as they transition to a drier state, and reswelling once they return to a hydrophilic state. This volume phase transition occurs when the local stimulus concentration crosses a critical value, about which the hydrophilicity of the polymer chains changes abruptly. These gels have been used to make surface crawlers or other pulsatile machines, but here we show that a very simple system comprising two oscillating gel spheres linked by a rigid rod can also swim in the inertialess Stokes regime -- albeit rather slowly. Developing a full continuum-mechanical model for a gel that employs a reaction described by the Brusselator model to generate oscillating chemical concentrations that couple to gel and fluid dynamics, we quantify the rate of shrinkage and swelling and associated flows as the gels pump out or draw in water. Herein, we derive analytical results for the swimming velocity of these swimmers, and upon placing them in a solute bath, identify two modes of behaviour upon encountering propagating reaction-diffusion waves: `bobbing' and `surfing'. Though somewhat slower than flagellated swimmers, the relative simplicity of the system, with no hinges or moving components, lends itself well to large scale production.
    \end{abstract}

    \maketitle

    \section{Introduction}
    There is great interest in the mechanics of swimming in viscous fluids at small scales where inertial effects can be neglected and the flow surrounding a swimmer is described by the (linear) Stokes equations \citep{lauga-2011-life_theorem}. Such is the environment faced by a variety of aquatic life such as zooplankton, individual cells including bacteria like \textit{E. coli}, and, within larger organisms, specialised motile cells such as spermatozoa \citep{lauga-2009_hydrodynamics_microorganisms}. All of these small bodies adopt a range of drag-propelled swimming strategies exploiting viscosity in their immediate surroundings, utilising flagella, cilia or even jets of fluid to propel themselves in a preferred direction. However, net motion through the fluid is impossible if the movement of the swimmer is reversible in time, a consequence of Purcell's famous `scallop theorem' \citep{purcell-1977_life_number} that results from the linearity of the equations governing low-Reynolds number flow. Thus, many of the familiar strategies for swimming in our intermediate-Reynolds number world (such as opening and closing the hinge of a scallop shell or idealised forms of breaststroke and front crawl swimming) are not available to microscopic life.

    The key to locomotion in Stokes flows is non-reciprocal actuation, where time-reversal symmetry no longer exists and net motion is permitted by the linear governing equations. In nature, this is achieved by means of rotary motors that take the form of helices or interact with boundaries to generate chiral flows producing translation (for example, the turning helix of \textit{E. coli} \citep{berg-2000_motile_bacteria,berg-2008_bacterial_motor}) or whip-like beating of flagella (like the tail of a spermatozoon \citep{suarez-2006_sperm_tract, lauga-2009_hydrodynamics_microorganisms, montenegro-johnson-2012_modelling_development}). Alternatively, flows can be generated by coordinated cilia on the surface of an organism -- described by the so-called `squirmer' models introduced by \citet{lighthill-1952_squirming_numbers} and \citet{blake-1971_spherical_propulsion}. All of these different approaches result in movement of varying efficiency.
    
    Learning from models seen in nature, a number of studies have designed artificial swimmers for microfluidics applications and microbots that can move through viscous fluids at small scales. The design of such swimmers began with \citet{purcell-1977_life_number} and the three-link swimmer, and has included designs that exploit connections that can change their length \citep{najafi-2004_simple_spheres,avron-2005_pushmepullyou_swimmer}, external actuation from magnetic fields \citep{box-2017_motion_flow} and non-reciprocity induced by background flows and/or non-Newtonian ambient fluids \citep{klotsa-2015_propulsion_swimmer,hewitt-2024_swimming_fluids}. Indeed, such devices have even been constructed using self-oscillating gels \citep{ahmed-2024_interactions_swimmers}, where chemical waves lead to peristaltic contractile behaviour in a hydrogel \citep{mao-2020_contraction_gels}, driving motion in one preferred direction. Alternatively, the repeated signal from an oscillating chemical reaction can drive non-reciprocal shape change that allows for swimming through a fluid via jet-driven propulsion \citep{tanasijevic-2022_jet_microgels}.

    One advantage that can be exploited in the design of synthetic swimmers is that we seek the simplest possible design to achieve non-reciprocal motion without the constraints imposed by what is biologically reasonable. In this vein, a class of swimmers built from chaining together spheres joined by rods of changing length has been developed over the years, starting with the famous three-sphere swimmer of \citet{najafi-2004_simple_spheres}. To achieve a non-reciprocal swimming pattern with identical spheres, it is necessary to join together three spheres using struts, since the inherent symmetry of a two-sphere configuration prevents net motion in a Newtonian fluid. Other authors have since considered two-sphere swimmers where non-reciprocity instead arises from deformation of one sphere but not of the other \citep{nasouri-2017_elastic_flow}, the behaviour of a complex fluid through which the swimmer travels \citep{montenegro-johnson-2012_physics_stokes}, or through imposing a background flow \citep{klotsa-2015_propulsion_swimmer}.
    
    The swelling and deswelling of responsive hydrogels in response to external stimuli makes them excellent candidates for the construction of non-reciprocal sphere-based swimmers. Instead of the length of the connection being varied, the sizes of the two spheres can be changed in a controllable and repeated pattern, leading to flows that result in net movement. A classic swimmer based on two spheres that change size due to transfer of internal (inviscid) fluid, the \textit{pushmepullyou} of \citet{avron-2005_pushmepullyou_swimmer} is known to be an efficient microswimmer that is realisable in experiments \citep{silverberg-2020_realization_numbers}. To achieve repeated deswelling--reswelling behaviour, it is possible to couple gel dynamics to an oscillating chemical reaction, usually the Belousov-Zhabotinsky (BZ) reaction \citep{belousov-1959_periodicheski_mekhanism}, to produce a so-called self-oscillating gel \citep{yoshida-1996_self_gel}, or, more specifically to the type of reaction, a BZ gel. These gels have been used in the past to achieve wormlike locomotion on a flat surface \citep{maeda-2007_self_gel} or the transport of material in persitaltic waves \citep{murase-2008_design_gel}, but in this example we use the simpler periodic deswelling cycle to give viscous drag--driven locomotion in the Stokes regime.
    
    In this paper, we introduce a continuum-mechanical model for self-oscillating gels that can comfortably explain the interplay between the numerous physical processes governing such pulsatile responsive hydrogels. We start by linking the deswelling and reswelling of a chemically-responsive gel to the local concentration of chemical species, explaining both how the equilibrium states in the presence and absence of chemical stimulus differ and how the gel reconfigures in response to its environment. From this modelling, we can derive the changes in shape of the hydrogel and the associated fluid flows. Finally, this framework is coupled with the Brusselator model \citep{glandsdorff-1971_thermodynamic_fluctuations} for an oscillating chemical reaction that describes the creation and depletion of stimulus chemical products, with a rate of reaction that can be influenced by composition of the gel in which it occurs. Together, this system exhibits rich dynamics beyond just a simple limit cycle, and we see that oscillations induced by such a reaction can lead to non-reciprocal swelling and drying behaviour, motivating the use of such a system to power locomotion in a low-Reynolds-number regime.

    In many simple geometries, the non-reciprocity of a self-oscillating gel is restricted to the shape change of the gel itself and fluid flows inside the pore scaffold: volume changes do not generate flows in the fluid bath surrounding the hydrogel since any bulk transport of water is exactly balanced by motion of the polymer chains. However, we introduce a simple spherically-symmetric geometry in which deswelling and reswelling of the gel produces a net flux of water, allowing for non-reciprocal fluid flows to be generated in response to simple pulsatile reaction forcing. These self-sustaining pumps are composed of a layer of responsive hydrogel coating an impermeable compressible core, and could therefore be realised straightforwardly in microfluidic applications.

    Using this modelling framework, we then design a two-sphere swimmer that exploits the viscous drag exerted on one sphere by the flows that are induced when the other swells or deswells to drive motion. With the archetypal non-reciprocal swimmer introduced by \citet{purcell-1977_life_number} as a model, we show how it is possible for two gels that deswell and reswell out of phase with one another to travel in one preferred direction over the course of a single stroke. This serves as a proof-of-concept for more physically realisable systems where the symmetry is broken not by out-of-phase responses to changes in chemical stimulus, which may be challenging to achieve in real systems, but by asymmetry in the rate of response across the two gels. The leading-order drift in all of these cases occurs at second order in the (small) ratio of the sphere rest radii to their separation, and is a slow drift on the background of an otherwise periodic motion, somewhat analogous in presentation to the second order Stokes drift \citep{stokes-1847_theory_waves} seen in transport due to water waves.

    In this example, we consider waves provided by pulsing concentrations of chemical species, arising from reaction--diffusion systems in which travelling waves of products are created. In section \ref{sec:surf_bob}, we investigate the interaction between these waves and the swimmers developed in this paper, showing how the locomotion of the swimmers falls into two distinct regimes: `bobbing' and `surfing'. In the former regime, as each chemical wave passes, the swimmer executes a single swimming stroke and travels forwards, though not as rapidly as the chemical wave, and so waits for the next wave to pass to `bob' forward again, much like a rubber duck being slowly transported forwards by passing water waves \citep{murase-2008_design_gel}. In the `surfing' regime, the swimmer begins to deswell as a chemical wave reaches it, but travels forwards faster than the wave, and therefore does not complete its stroke. The wave then catches up and the process repeats, allowing the swimmer to `surf' at the same speed as the propagating chemical front, potentially much faster than the second-order drift usually experienced by such swimmers. To drive these behaviours, wavelike chemical concentration profiles from oscillating reactions are used, that are well-attested in the experimental literature \citep{field-1985_oscillations_systems,mao-2020_contraction_gels}, with wavefronts of high concentration followed by regions of low concentration that travel through space.

    \section{Modelling oscillating gels}
    \label{sec:bzgel}
    To design a swimming device that employs swelling and deswelling of responsive gels to drive motion, we must provide some repeated external stimulus to actuate these swell-deswell strokes. In other studies, this can be achieved by periodic heating and cooling \citep{tanasijevic-2022_jet_microgels}, but here we focus our attention on (somewhat) self-sustaining stimuli, in the form of oscillating chemical reactions coupled to chemically-responsive hydrogels. In such reactions, the concentration of products cycles repeatedly over a fixed, predictable, time period. The first oscillating chemical reaction to be demonstrated experimentally was the Belousov-Zhabotinsky (BZ) reaction, discovered by \citet{belousov-1959_periodicheski_mekhanism}, where the oscillatory mechanism results from autocatalytic behaviour (i.e. the products catalyse the reaction itself).

    Coupling BZ reactions to chemically-responsive gels in order to drive periodic swelling and deswelling behaviour is not without precedent -- the study of so-called BZ gels began with the experimental investigation of \citet{yoshida-1996_self_gel}, who demonstrated the feasibility of these systems and proposed a number of uses ranging from periodic drug release to pacemaking and crawling-type locomotion. A number of more recent studies have demonstrated applications in this latter category, including in the creation of gels that locomote due to periodic travelling chemical waves within their structure \citep{ren-2020_chemomechanical_signals}, gels that crawl along solid surfaces, again as a result of internal peristaltic waves \citep{maeda-2007_self_gel}, and those that attract or repel one another \citep{epstein-2012_chemical_media} to drive collective motion. There are also a number of modelling approaches that aim to accurately describe the shape-changing response of gels to changing chemical concentration fields \citep{yashin-2006_modeling_reaction, yashin-2007_theoretical_gels, kuksenok-2008_three_reaction}. These models are often based on a lattice of springs that can be generalised to arbitrary geometries, making them especially suited for rapid numerical solutions, but at times disguising the physical processes underlying the swelling and drying of the gel, and making coupling to fluid flows more challenging.

    In the present study, we seek a continuum-mechanical description of the behaviour of a self-oscillating gel that depends on only a small number of macroscopic parameters, and does not require a complicated parametrisation of the micro-scale interactions between solute and polymer molecules to understand the deswelling and reswelling process. Following \citet{webber-2025_poromechanical_pumps}, we couple reaction and gel dynamics through a chemical-concentration-dependent osmotic pressure, and can therefore easily use force balances to determine the evolution of polymer fraction in time and the associated fluid flows. In addition to the broad applicability of a model independent of specific polymer--solute interaction mechanisms \citep{kuksenok-2013_chemo_communication}, this will also allow us to describe these gels using only three coupled partial differential equations for composition and concentration of chemical species.

    \subsection{Responsive gel modelling}
    In a recent study \citep{webber-2025_oscillating_hydrogels}, we used the linear-elastic-nonlinear-swelling (LENS) model for responsive hydrogels \citep{webber-2023_linear_gels,webber-2025_poromechanical_pumps} to describe the deswelling and reswelling response of a hydrogel in the form of a coupled series of partial differential equations that allow us to understand the interplay between reaction dynamics, elastic deformation of the hydrogel and pumping of interstitial (pore) fluid. The chemical reaction was described using the Brusselator model, which gives qualitatively similar kinetics to reactions such as the BZ reaction. This is the same approach that we utilise in the present study, modelling a hydrogel with a rest-state polymer volume fraction that changes sharply as a critical concentration of chemical species $Y$, $Y_c$, is exceeded. 
    
    In the absence of species $Y$, the gel is in a swollen state, and balancing osmotic pressures with elastic stresses that arise from isotropic deformation of the polymer scaffold, reaches a uniform polymer fraction $\phi_{00}$. The form of the osmotic pressure changes when $Y > Y_c$, and the same balance of pressures results in a new equilibrium polymer fraction $\phi_{0\infty}$ that corresponds to a deswollen gel. In the transient states between these two extremes, fluid flow is driven from areas of high to low pressure through the pore spaces of the gel, and swelling and deswelling are governed by these interstitial flows $\boldsymbol{u}$, measured relative to a deforming polymer scaffold, with
    \begin{equation}
        \boldsymbol{u} = \frac{D(\phi)}{\phi}\bnabla \phi \qq{where} D(\phi) = \frac{k(\phi)}{\mu_l}\left[\phi\pdv{\Pi}{\phi} + \frac{4 \mu_s(\phi)}{3}\left(\frac{\phi}{\phi_{00}}\right)^{1/3}\right].
        \label{eqn:interstitial_u}
    \end{equation}
    Here, $k(\phi)$ is the hydrodynamic permeability of the hydrogel, $\Pi(\phi)$ is the osmotic pressure and $\mu_s(\phi)$ is the shear modulus \citep{webber-2023_linear_gels}. Flow is governed by the viscosity of the interstitial water, $\mu_l$. Broadly, this result can be interpreted as showing that fluid flows from more swollen to drier regions of a hydrogel, as would be expected, with the rate of flow modulated by the local permeability and driven by how far from an equilibrium state that the gel lies. Eventually, a steady state will be reached where imposed elastic stresses on the gel will be balanced by osmotic pressures, and in the absence of such stresses a uniform equilibrium state with zero osmotic pressure will be approached.
    
    For simplicity, we will restrict our attention in this study to cases where $k$ and $\mu_s$ are constants independent of $\phi$ and where $\Pi(\phi) = \Pi_0[\phi-\phi_0(Y)]/\phi_0(Y)$, with $\phi_0$ the equilibrium polymer fraction (i.e. $\phi_0 = \phi_{00}$ below the critical chemical concentration and $\phi_{0\infty}$ above it), which capture the key qualitative behaviours we seek to model without forcing us to choose a specific constitutive model. Somewhat surprisingly, the value of $\mu_s$ often depends only weakly on swelling state; indeed, it is independent of $\phi$ when a Flory-Rehner model is coupled to the assumptions of the LENS model \citep{webber-2025_poromechanical_pumps}, justifying our assumption of a constant value. Though there are some qualitative differences in interstitial fluxes when $k$ depends on polymer fraction \citep{macminn-2016_large_material}, since flow is constricted as a gel deswells, as discussed in appendix \ref{app:permeability}, we limit ourselves to the simplest possible model with constant $k$ in the present work, and can add the feedback effect of deswelling and reswelling later. 
    
    Conservation of water and polymer, alongside the assumption that both phases are (separately) incompressible, therefore give the evolution equation
    \begin{equation}
        \pdv{\phi}{t} + \boldsymbol{q}\bcdot\bnabla \phi = \bnabla \bcdot \left[D(\phi)\bnabla\phi\right] \qq{where} D(\phi) = \frac{k\Pi_0}{\mu_l} \times \begin{dcases}\frac{\phi}{\phi_{00}} + \frac{4\mu_s}{3\Pi_0}\left(\frac{\phi}{\phi_{00}}\right)^{1/3} & Y \le Y_c \\ \frac{\phi}{\phi_{0\infty}} + \frac{4\mu_s}{3\Pi_0}\left(\frac{\phi}{\phi_{00}}\right)^{1/3} & Y > Y_c\end{dcases},
        \label{eqn:lens}
    \end{equation}
    where $\boldsymbol{q}=\boldsymbol{u}+\boldsymbol{u_p}$ is the phase-averaged flux vector comprising contributions from volumetric fluid flux $\boldsymbol{u}$ and polymer velocity $\boldsymbol{u_p}$. Unlike the fluid flux $\boldsymbol{u}$, this quantity represents a flux of gel -- both polymer and water phases -- and is measured in the lab frame and not relative to a deforming background. In \citet{webber-2023_linear_gels}, it was shown that $\bnabla\bcdot\boldsymbol{q} = 0$, a consequence of the fact that the volume of gel can only change by the addition or removal of water. Alongside boundary conditions on fluid flow, pressure and stress and an expression for how the shape of the gel changes (found either via conservation of polymer in simple geometries with one degree of freedom, or through more complicated means when free to swell or dry in multiple directions \citep{webber-2023_linear_formulation}), this allows us to find the transient response of a hydrogel to a change in chemical stimulus.

    \subsection{Reaction dynamics and solute transport}
    In order to understand how the chemical concentration of the stimulus species changes in time as a result of an oscillating chemical reaction, we must prescribe a reaction model to describe the creation and depletion of this species. The BZ reaction is the most well-known oscillating chemical reaction, but is also very complex, with a mechanism involving many steps. This mechanism has been studied extensively in the literature, starting with the mechanism proposed by Field, K\"or\"os and Noyes \citep{field-1972_oscillations_system}, and its later simplification, the Oregonator \citep{field-1985_oscillations_systems}. Even this latter simplified model involves five reaction steps and an undetermined stoichiometric constant, making mathematical analysis of the reaction mechanism challenging. Prior to the introduction of the Oregonator model, Prigogine and Lefever \citep{prigogine-1968_symmetry_systems} introduced the much simpler Brusselator model for autocatalytic oscillating reactions, featuring four reaction steps. This model qualitatively reproduces the oscillating behaviour seen in the BZ reaction, and even though it is phenomenological in nature, it has previously been used successfully in the modelling of self-oscillating gels \citep{borckmans-2003_model_gels} and could be replaced in the analysis that follows with one of the aforementioned reaction-specific approaches if quantitative agreement is necessary.

    We start by considering the reaction dynamics in a pure fluid bath, using the Brusselator reaction model \citep{glandsdorff-1971_thermodynamic_fluctuations}, where the four reaction steps have corresponding rate constants $r_1,\,\dots,\,r_4$:
    \begin{equation}
        A \xrightarrow[r_1]{} X, \quad 2 X + Y \xrightarrow[r_2]{} 3X, \quad B + X \xrightarrow[r_3]{} Y + F \qq{and} X \xrightarrow[r_4]{} G.
    \end{equation}
    Thus, the concentrations of species $A$, $B$, $F$, $G$, $X$ and $Y$ evolve following the coupled system of equations
    \begin{align}
        \dv{[A]}{t} &= -r_1[A],\quad \dv{[B]}{t} = -r_3 [B][X],\quad \dv{[F]}{t} = r_3[B] [X],\quad \dv{[G]}{t} = r_4 [X],\notag \\ \dv{[X]}{t} &= k_1[A] + k_2 [X]^2[Y] -k_3 [B] [X] - k_4 [X] \qq{and} \dv{[Y]}{t} = -k_2 [X]^2[Y] + k_3 [B][X].
        \label{eqn:brusselator_raw}
    \end{align}
    At this juncture, we note that the concentration of $F$ and $G$ are unimportant for the dynamics of the reaction itself, and therefore we neglect the influence of these species on the system. Furthermore, we take $A$ and $B$ to be in excess, so assume that they are not noticeably depleted by the first two equations of the system, and can be treated as constants. Both of these assumptions are standard in the analysis of such reactions \citep{prigogine-1968_symmetry_systems}, so we now rescale the concentrations\footnote{Equation \eqref{eqn:brusselator_raw} can be rescaled by taking $\tilde{A} = r_1 r_2^{1/2} [A]/ r_4^{3/2}$, $\tilde{B} = r_3 [B] / r_4$, $\tilde{X} = \sqrt{r_4/r_2}[X]$ and $\tilde{Y} = \sqrt{r_4/r_2}[Y]$.} and find that the reaction can be described by the second-order system
    \begin{equation}
        \dv{\tilde{X}}{t} = r_4\left[\tilde{A} + \tilde{X}^2 \tilde{Y} - (1+\tilde{B})\tilde{X}\right] \qq{and} \dv{\tilde{Y}}{t} = r_4\left[- \tilde{X}^2\tilde{Y} + \tilde{B}\tilde{X}\right].
        \label{eqn:brusselator_rescaled}
    \end{equation}
    This system admits a steady-state solution $(\tilde{X},\,\tilde{Y}) = (\tilde{A},\,\tilde{B}/\tilde{A})$, which is unstable in cases $\tilde{B}>1+\tilde{A}^2$, giving limit-cycle dynamics in these situations via a Hopf bifurcation as the fixed point becomes unstable \citep{lefever-1971_chemical_oscillations}. In reality, these periodic solutions will decay in time, but the assumption that $A$ and $B$ cannot be used up removes this effect, as borne out in experiments which show very slow decay \citep{yoshida-1996_self_gel}.

    In a self-oscillating gel, reactions occur inside the gel scaffold, with a pore fluid fraction $1-\phi$. Per unit volume, there is therefore a quantity $(1-\phi)c$ of a species with concentration $c$. The local concentration of solute species can change by diffusion of solute within the pore fluid or by advection along with the flow of interstitial fluid through the matrix. We define a solute flux per unit volume $\boldsymbol{j}_c$ and find an evolution equation for the amount of solute per unit volume of hydrogel,
    \begin{equation}
        \boldsymbol{j}_c = \boldsymbol{u}c - (1-\phi)D_m\bnabla c \qq{hence} \pdv{t}\left[(1-\phi)c\right] + \bnabla \bcdot \left(\boldsymbol{u}c\right) = \bnabla \bcdot \left[(1-\phi) D_m \bnabla c\right],
        \label{eqn:gel_transport}
    \end{equation}
    where $D_m$ is the molecular diffusion coefficient and $\boldsymbol{u}$ is the volume flux of water through the hydrogel driven by gradients in pore pressure. For the modelling that follows, we assume that molecular diffusivity is a constant in space, and that there are no effects such as dispersion that will have an effect on transport in the pore spaces, justified by noting that fluid flows in gels are slow \citep{webber-2025_poromechanical_pumps}. Outside of the hydrogel, where $\boldsymbol{u}$ now describes an incompressible fluid flow,
    \begin{equation}
        \boldsymbol{j}_c = \boldsymbol{u}c - D_m\bnabla c \qq{so} \pdv{c}{t} + \boldsymbol{u}\bcdot\bnabla c = D_m \nabla^2 c.
    \end{equation}

    Finally, there exists a coupling between the degree of deswelling in a hydrogel and the local reaction rate. In order to allow mechanical processes like deformation to feed back into the dynamics of the chemical oscillation, catalysts for the reaction are often chemically bonded to the cross-linked polymer chains forming the scaffold of the hydrogel \citep{yoshida-1996_self_gel}, and hence their local concentration is proportional to $\phi$. As the gel deswells, the chains pack tighter and the concentration of catalyst per unit volume increases, leading to a higher rate of reaction, with the reverse process leading to lower reaction rates in a swollen gel. Hence, we replace the rate constant $r_4$ in equation \eqref{eqn:brusselator_rescaled} with $r_4 \phi^{\alpha}$, where $\alpha$ is a stoichiometric constant for the catalytic process. Together, equations \eqref{eqn:brusselator_rescaled} and \eqref{eqn:gel_transport} combine to give
    \begin{subequations}
        \begin{align}
            \pdv{t}\left[(1-\phi)X\right] &+ \bnabla\bcdot\left[\frac{D(\phi)X}{\phi}\bnabla\phi\right] = D_m\bnabla\bcdot\left[(1-\phi)\bnabla X\right] + r_4(1-\phi)\phi^\alpha\left[A + X^2 Y - (1+B) X\right] \quad \text{and} \\
            \pdv{t}\left[(1-\phi)Y\right] &+ \bnabla\bcdot\left[\frac{D(\phi)Y}{\phi}\bnabla\phi\right] = D_m\bnabla\bcdot\left[(1-\phi)\bnabla Y\right] + r_4(1-\phi)\phi^\alpha\left[-X^2 Y + BX\right],
        \end{align}
        \label{eqn:react_diffuse}%
    \end{subequations}
    within the gel, dropping tildes for legibility and using equation \eqref{eqn:interstitial_u} for the interstitial fluid flux. We assume that the reaction is confined only to the gel itself, since there are no catalyst molecules present in the surrounding fluid. Later in this study we will consider the opposite case, where no reaction takes place in the gel but it responds to changes in chemical concentration arising from reactions in a surrounding bath of fluid. Of course, equations \eqref{eqn:react_diffuse} only describe chemical species $X$ and $Y$ and do not consider $A$ and $B$, but since these are not created or destroyed in reactions and are initially spatially-uniformly distributed and in excess, the concentration per unit pore fluid does not change and these concentrations are just parameters for the reaction model.

    \subsection{Non-dimensionalisation and a model problem}
    \label{sec:bz_gel_nd}
    \begin{figure}
        \centering
        \includegraphics[width=0.5\linewidth]{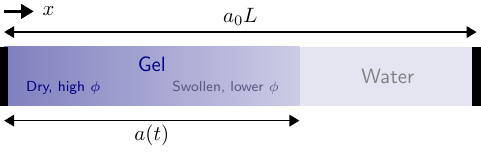}
        \caption{A diagram illustrating the one-dimensional model problem of a self-oscillating gel placed in a container of water with walls at $x=0$ and $x=a_0 L$, where reactions take place only within the gel, but solute species involved in the reaction can diffuse into the water.}
        \label{fig:schematic_experiment}
    \end{figure}
    For a gel of characteristic lengthscale $a_0$, equation \eqref{eqn:lens} shows that interstitial fluid flows reconfigure the gel on a characteristic timescale $t_{\text{pore}} = \mu_l a_0^2/k\Pi_0$ set by balancing the viscous resistance to flow and the strength of osmotic pressures. The coupled gel--reaction problem also has a reaction timescale, $t_{\text{react}} = 1/k_4 \phi_{00}^\alpha$ and a diffusive transport timescale $t_{\text{diff}} = a_0^2/D_m$, all of which describe the rate at which different processes contribute to the dynamics of the system. We scale all lengths by $a_0$ and all times by $t_{\text{pore}}$, letting $\tau=t/t_{\text{pore}}$, and note that the gel evolution equation \eqref{eqn:lens} becomes
    \begin{equation}
        \pdv{\Phi}{\tau} + \bar{\boldsymbol{q}}\bcdot\bar{\bnabla}\Phi = \bar{\bnabla}\bcdot\left[\left(\frac{\Phi}{\Phi_0(Y)} + \frac{4\mathcal{M}}{3}\Phi^{1/3}\right)\bar\bnabla\Phi\right] \qq{with} \Phi_0(Y) = \begin{dcases}1 & Y \le Y_c \\ \phi_{0\infty}/\phi_{00} & Y > Y_c\end{dcases},
        \label{eqn:sys_pf_evol}
    \end{equation}
    where $\Phi = \phi/\phi_{00}$, $\bar{\boldsymbol{q}}$ is the phase-averaged flux rescaled by $a_0/t_{\text{pore}}$ and $\bar{\bnabla}$ denotes the gradient taken with respect to spatial coordinates scaled with $a_0$. The parameter $\mathcal{M}=\mu_s/\Pi_0$ quantifies the relative importance of shear stresses over osmotic pressures.
    
    Making the same scalings in equations \eqref{eqn:react_diffuse}, alongside the distinguished limit $\phi_{00} \ll 1$ (such that diffusion in the gel pores is similar to that in pure water), results in the non-dimensional transport equations
    \begin{subequations}
        \begin{align}
            \pdv{X}{\tau} &+ \bar{\bnabla}\bcdot\left[\left(\frac{1}{\Phi_0(Y)} + \frac{4\mathcal{M}}{3}\Phi^{-\frac{2}{3}}\right)X\bar{\bnabla}\Phi\right] = \mathcal{D}\bar{\nabla}^2 X + Da\Phi^\alpha\left[A + X^2 Y - (1+B) X\right]\!,\!\! \\
            \pdv{Y}{\tau} &+ \bar{\bnabla}\bcdot\left[\left(\frac{1}{\Phi_0(Y)} + \frac{4\mathcal{M}}{3}\Phi^{-\frac{2}{3}}\right)Y\bar{\bnabla}\Phi\right] = \mathcal{D}\bar{\nabla}^2 Y + Da\Phi^\alpha\left[-X^2 Y + BX\right],
        \end{align}
        \label{eqn:sys_react_diffuse}%
    \end{subequations}
    in the gel and
    \begin{equation}
        \pdv{\tau}\begin{pmatrix}X\\Y\end{pmatrix} + \bar{\boldsymbol{V}}\bcdot\bar{\bnabla}\begin{pmatrix}X\\Y\end{pmatrix} = \mathcal{D} \bar{\nabla}^2 \begin{pmatrix}X\\Y\end{pmatrix},
        \label{eqn:sys_diffuse}
    \end{equation}
    in the external fluid, where $\bar{\boldsymbol{V}}$ represents the fluid velocity scaled by $a_0/t_{\text{pore}}$. The contributions from diffusion and the reaction are quantified by the diffusion parameter $\mathcal{D} = t_{\text{pore}}/t_{\text{diff}}$ and the Damk\"ohler number $Da = t_{\text{pore}}/t_{\text{react}}$, respectively, allowing us to describe the separation between the three timescales in this problem. This system, comprising equations \eqref{eqn:sys_pf_evol}, \eqref{eqn:sys_react_diffuse} and \eqref{eqn:sys_diffuse} can be solved alongside boundary conditions and polymer conservation constraints on the gel to find its extent and deduce the evolution of any self-oscillating gel we wish to model.

    To illustrate the utility of this approach, we will consider a one-dimensional problem of a responsive hydrogel in a closed box $0 \le x \le a_0 L$ ($L>1$), pictured in figure \ref{fig:schematic_experiment}. Such one-dimensional self-oscillating gels have been modelled in the past, both using the Brusselator model \citep{borckmans-2003_model_gels}, the Oregonator \citep{yashin-2006_modeling_reaction}, and even non-oscillating reactions coupled with mechanical feedback that drives a loop \citep{boissonnade-2009_oscillatory_model}. When fully swollen, the gel occupies the region $0 \le x \le a_0$, and it remains anchored to the boundary at $x=0$ as it deswells and reswells, occupying the space $0 \le x \le a(t)$ at time $t$. At the left-hand boundary, there is no diffusive flux of chemical species $X$ and $Y$, so
    \begin{equation}
        \pdv{X}{\bar{x}} = \pdv{Y}{\bar{y}} = 0 \quad \text{on $\bar{x}=0$.}
    \end{equation}
    Furthermore, the boundary is impenetrable to fluid, so $\pdv*{\phi}{\bar{x}} = 0$ here as well, using equation \eqref{eqn:interstitial_u} for the interstitial fluid velocity. Since the polymer velocity must also be zero at this boundary, the phase-averaged material flux $\bar{q}=0$ on $\bar{x}=0$, and conservation of mass implies $\pdv*{\bar{q}}{\bar{x}}=0$ \citep{webber-2023_linear_gels}, hence $\bar{q} \equiv 0$ in this geometry.
    
    It is clear that the local concentration (in water) of species $X$ and $Y$ must be the same either side of the gel--water interface, even though the true volumetric concentration will change owing to the presence of a polymer phase. Hence, we take $X$ and $Y$ continuous at $x=\bar{a}(\tau)$. We assume there to be no normal stresses at the gel--water interface so it is always swollen to the equilibrium value $\phi_0(Y)$, since the fluid pressure is continuous across phases, there is no deviatoric strain, and thus osmotic pressure must be zero to balance stresses. This supplies a Dirichlet boundary condition on $\phi$ and closes the system of equations for gel composition.

    At a gel--water interface, the normal fluid velocity is specified through a mass conservation argument, taking into account the moving boundary \citep{xu-2022_comparison_interface}, so that
    \begin{equation}
        \bar{\boldsymbol{V}}\bcdot\boldsymbol{n} = \bar{\boldsymbol{q}}\bcdot\boldsymbol{n},
    \end{equation}
    In our case, since the total fluid flux is identically zero in the gel, the swelling and deswelling does not generate a flow field in the water bath. In order to increase its volume, the gel simply `engulfs' fluid that occupies the space it expands into, and to deswell, the polymer scaffold draws back through the fluid, leaving water in place. In spite of this, there are still fluid flows from more swollen to less swollen regions of gel within the pore structure, since water is driven by differences in stress through the matrix. Finally, the gel thickness is set by polymer conservation, with
    \begin{equation}
        \int_0^{a(t)}{\phi\,\mathrm{d}x} = \phi_{00}a_0.
        \label{eqn:mass_cons}
    \end{equation}
    Then, non-dimensionalising with the same scalings as introduced above,
    \begin{subequations}
        \begin{gather}
            \pdv{\Phi}{\tau} = \pdv{\bar{x}}\left[\left(\frac{\Phi}{\Phi_0(Y)} + \frac{4\mathcal{M}}{3}\Phi^{1/3}\right)\pdv{\Phi}{\bar{x}}\right] \qq{with} \begin{dcases}\pdv*{\Phi}{\bar{x}} = 0 & \text{on $\bar{x}=0$,} \\ \Phi = \Phi_0(Y) & \text{on $\bar{x}=\bar{a}(\tau)$}\end{dcases}, \\
            \pdv{X}{\tau} + \pdv{\bar{x}}\left[\left(\frac{1}{\Phi_0(Y)} + \frac{4\mathcal{M}}{3}\Phi^{-\frac{2}{3}}\right)X\pdv{\Phi}{\bar{x}}\right] = \mathcal{D}\pdv[2]{X}{\bar{x}} + Da\Phi^\alpha\left[A + X^2 Y - (1+B) X\right], \\
            \pdv{Y}{\tau} + \pdv{\bar{x}}\left[\left(\frac{1}{\Phi_0(Y)} + \frac{4\mathcal{M}}{3}\Phi^{-\frac{2}{3}}\right)Y\pdv{\Phi}{\bar{x}}\right] = \mathcal{D}\pdv[2]{Y}{\bar{x}} + Da\Phi^\alpha\left[-X^2 Y + BX\right], \intertext{in $0 \le \bar{x} \le \bar{a}(\tau)$ and with $\pdv*{X}{\bar{x}}=\pdv*{Y}{\bar{Y}}=0$ on $\bar{x}=0$, and}
            \centering\pdv{\tau}\begin{pmatrix}X\\Y\end{pmatrix} = \mathcal{D} \pdv[2]{\bar{x}} \begin{pmatrix}X\\Y\end{pmatrix},
        \end{gather}
        \label{eqn:full_1d_problem}%
    \end{subequations}
    in $\bar{a}(\tau) \le \bar{x} \le L$ with $\pdv*{X}{\bar{x}}=\pdv*{Y}{\bar{x}}=0$ on $\bar{x}=L$. The extent $\bar{a}(\tau)$ is set implicitly through a rescaled version of the integral condition \eqref{eqn:mass_cons}, and both $X$ and $Y$ are continuous at this interface.

    \begin{figure}
        \centering
        \includegraphics[width=0.9\linewidth]{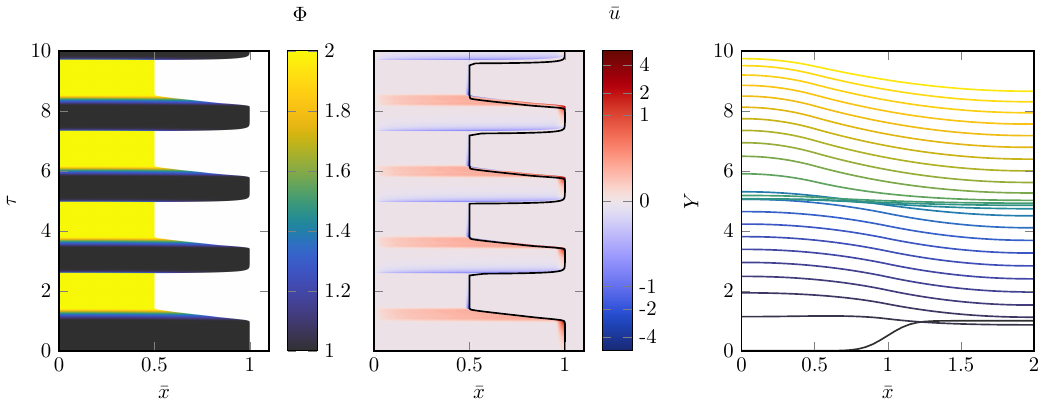}
        \caption{Plots of a solution to the one-dimensional Brusselator gel problem described by equations \eqref{eqn:full_1d_problem} with $\alpha = 1$, $\mathcal{M}=1$, $\mathcal{D}=1$, $Da=10$ and $L=2$, as well as $A=1$ and $B=5$, with the gel initially at rest and $X\equiv0$, $Y=1$ outside the gel and $Y=0$ inside. The gel fully deswells to $\Phi \equiv \phi_{0\infty}/\phi_{00}=2$ before reswelling cyclically, as shown in the first panel. In the second panel, the fluid velocity field is plotted, showing flows towards the interface ($\bar{u}>0$) when deswelling and away from it as the hydrogel equilibrates. Finally, the concentration of species $Y$ is plotted at a number of time intervals during the deswelling process, showing both how it diffuses throughout the entire system and increases owing to the reaction, plateauing at $Y\approx Y_c = 5$ as the gel deswells and then increasing further.}
        \label{fig:1d_results}
    \end{figure}
    In figure \ref{fig:1d_results}, the time evolution of the composition of the hydrogel and the driving chemical concentration field is plotted, showing how the concentration of chemical species $Y$ is highest near $\bar{x}=0$, where chemical species created in the reaction have no chance to diffuse out of the gel. Eventually, when $Y$ crosses the threshold $Y_c$ at $\bar{x}=1$, the gel begins to deswell, and fluid (containing solute chemicals) is transported into the bath, momentarily lowering the concentration of $Y$ inside the gel and slowing the deswelling. After this, the deswelling restarts, and the gel completely shrinks before reswelling when the chemical concentration again drops below $Y_c$. It is clear that variation of $Y$ within the gel has little effect on the overall qualitative dynamics, because the swelling and deswelling only triggers when $Y$ crosses $Y_c$ at the interface with only a slight modification of the internal diffusivity if this happens within the gel.

    \begin{figure}
        \centering
        \includegraphics[width=0.9\linewidth]{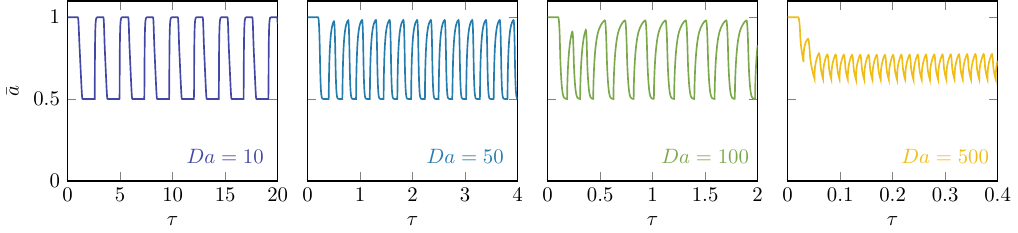}
        \caption{Plots of the gel thickness $\bar{a}(\tau)$ for different values of the Damk\"ohler number $Da$, showing that, when gel response times are faster (low $Da$) relative to the rate of reaction, the hydrogel will deswell and reswell in full in response to the changing chemical concentration, whilst there is incomplete swelling and deswelling for gels where the reaction happens more quickly on the timescale of polymer scaffold reconfiguration. Besides the Damk\"ohler number, the same parameters as in figure \ref{fig:1d_results} are used here.}
        \label{fig:vary_da}
    \end{figure}
    In figure \ref{fig:vary_da}, the effect of changing the Damk\"ohler number whilst holding the other parameters fixed is illustrated, showing that increasing the reaction rate relative to the poroelastic adjustment rate leads eventually to such fast oscillations in chemical concentration that the gel cannot fully deswell or reswell in response, and its thickness instead performs smaller-amplitude oscillations around a partially-deswollen state. Unlike our recent work \citep{webber-2025_oscillating_hydrogels} that assumes the adjustment of a hydrogel to changes in its surroundings is instantaneous on the timescale of the reaction ($Da=0$), richer dynamics are seen here, with a time lag between the chemical concentration crossing the threshold and a subsequent reconfiguration of the gel modifying the chemical kinetics.

    This figure shows clearly that the impact of spatial heterogeneity is minimised in the low-$Da$ limit, wherein the mechanical response to the changing chemical reaction field is near-instantaneous and depends only on the value of $Y$ at $\bar{a}(\tau)$, with no lag. However, at larger Damk\"ohler numbers where the gel's reconfiguration happens much slower than the reaction kinetics, internal transport of fluid shows that the interfacial concentration of $Y$ can no longer be used as a proxy for the mechanical state of the gel. 
    
    It is also apparent that the nature of deswelling and reswelling is entirely different -- in figure \ref{fig:1d_results}, deswelling is seen to be much slower than reswelling, whilst the reverse is clear for larger Damk\"ohler numbers in figure \ref{fig:vary_da}. These results show very clearly how self-oscillating gels can convert a totally `symmetric' periodic pulse into an asymmetric response, apparent in nonreciprocal changes in their shape. This asymmetry in swelling and deswelling behaviour can be exploited to achieve locomotion when in contact with solid surfaces (for example in squirmers and crawlers). However, since flows are not generated in the surrounding fluid, as a result of the fact that $q \equiv 0$ in this specific example, this nonreciprocal actuation does not extend to the surroundings of the gel and cannot be used to propel these gels through an entirely fluid medium, unless we conceive of a geometry in which there is net flux of gel material.

    \section{Generating fluid flows from responsive gels}
    \label{sec:coreshell}
    In order to design swimmers that exploit the periodic forcing supplied by an oscillating chemical reaction, we seek geometrically simple responsive gel components that pump fluid into their surroundings and thus generate fluid flows that oscillate as deswelling or reswelling occurs. It is tempting to assume that a simple responsive gel sphere will generate an outwards radial flux as it deswells (and, correspondingly, an inwards flux as it reswells), but in fact the phase-averaged material flux $q$ is everywhere zero in such gels \citep{webber-2023_linear_gels} and so any outward fluid flows are counteracted by the moving boundary and a need to draw water in to areas that were once occupied by gel. These two effects exactly balance one another out, and the radial fluid velocity on the gel--water interface can be shown to be exactly zero, much as in the one-dimensional case above. 
    
    To overcome this shortcoming, consider a gel layer $\rho_1(t)\le r \le \rho_2(t)$ coating the surface of an impermeable compressible sphere $0 \le r \le \rho_1(t)$. The initial state is of a gel of polymer fraction $\phi_{00}$ occupying the space $\rho_{10} \le r \le \rho_{20}$, and the subsequent evolution of polymer fraction is assumed to be spherically symmetric and described by
    \begin{equation}
        \pdv{\phi}{t} + q \pdv{\phi}{r} = \frac{1}{r^2}\pdv{r}\left[r^2 D(\phi)\pdv{\phi}{r}\right],
    \end{equation}
    where $D(\phi)$ is the polymer diffusivity of equation \eqref{eqn:interstitial_u} and $q$ is a phase-averaged material flux. Since this flux is solenoidal, $r^2 q$ must be a constant in space, so $q=K/r^2$ for some constant $K$. In a sphere, regularity at the origin forces this constant to be zero, but in this spherical shell geometry, there must be a nonzero flux of material on the inner surface, forcing $K$ to be nonzero. Introduce standard non-dimensionalisations using $\rho_{20}$ as the characteristic lengthscale,
    \begin{equation}
        \tau = \frac{t}{t_{\text{pore}}}, \quad R = \frac{r}{\rho_{20}}, \quad R_1 = \frac{\rho_1}{\rho_{20}}, \quad R_2 = \frac{\rho_2}{\rho_{20}}, \quad \lambda = \frac{\rho_{10}}{\rho_{20}} \qq{and} \mathcal{K} = \frac{t_{\text{pore}}}{\rho_{20}^3}K.
    \end{equation}
    Then, the evolution of scaled polymer fraction $\Phi = \phi/\phi_{00}$ is described by
    \begin{equation}
        \pdv{\Phi}{\tau} + \frac{\mathcal{K}}{R^2} \pdv{\Phi}{R} = \frac{1}{R^2}\pdv{R}\left[R^2\left(\frac{\Phi}{\Phi_0} + \frac{4\mathcal{M}}{3}\Phi^{1/3}\right)\pdv{\Phi}{R}\right] \qq{with} 4\pi\int_{R_1}^{R_2}{R^2\Phi\,\mathrm{d}R} = \frac{4\pi}{3}\left(1-\lambda^3\right),
        \label{eqn:evolution_coreshell}
    \end{equation}
    for $\Phi_0$ the equilibrium polymer fraction as defined in equation \eqref{eqn:sys_pf_evol}, and with the integral constraint enforcing conservation of polymer. We are able to decouple gel dynamics from the chemical reaction problem that changes the value of $\Phi_0$ if we work in the low Damk\"ohler number limit of the previous section, and assume that the gel responds fully to a change in chemical stimulus before this chemical stimulus changes.
    
    The governing equation \eqref{eqn:evolution_coreshell} is to be solved subject to $\pdv*{\Phi}{R} = 0$ on $R=R_1(\tau)$ since this boundary is impermeable and equation \eqref{eqn:interstitial_u} shows how fluid fluxes are proportional to gradients in $\phi$. On the outer radius $R=R_2(\tau)$, we take the pore pressure to be continuous across the gel--water interface and therefore osmotic pressures must balance deviatoric strains, as seen at the interface between a swelling sphere and water in \citet{webber-2023_linear_gels}, so that
    \begin{equation}
        \Phi-\Phi_0 = 2\mathcal{M}\Phi_{0}\epsilon_{rr}.
        \label{eqn:interfacial_bc_raw}
    \end{equation}
    The dimensionless radial displacement from fully-swollen equilibrium, $\xi$ (scaled with $\rho_{10}$) can be computed from the polymer fraction using the divergence of the displacement field \citep{webber-2023_linear_gels}, so
    \begin{equation}
        \pdv{\xi}{R} + 2\frac{\xi}{R} = 3\left(1-\Phi^{1/3}\right) \qq{so} \xi = \frac{3}{R^2}\int_{R_1}^{R}{u^2(1-\Phi^{1/3})\mathrm{d}u} + \frac{R_1^2(R_1-\lambda)}{R^2},
    \end{equation}
    since $\xi = R_1-\lambda$ on $R=R_1$. Hence, since the radial deviatoric strain is defined by $\epsilon_{rr} = \pdv*{\xi}{R} - 1 + \Phi^{1/3}$, the condition of equation \eqref{eqn:interfacial_bc_raw} can be restated, using
    \begin{equation}
        \left.\pdv{\xi}{R}\right|_{R=R_2} = 3\left(1-\Phi^{1/3}\right) - \frac{2(R_2 - 1)}{R_2} \qq{thus} \Phi-\Phi_0 = 4\mathcal{M}\Phi_{0}\left(R_2^{-1} - \Phi^{1/3}\right).
        \label{eqn:interfacial_bc}
    \end{equation}
    These two boundary conditions are not alone sufficient to determine the evolution of polymer fraction; we must also deduce the position of the two radial interfaces and the value of the constant $\mathcal{K}$. On the innermost surface $R=R_1$, the interstitial fluid flux is zero relative to the reconfiguring polymer scaffold, since the boundary is impermeable, and therefore $q$ is simply equal to the polymer velocity. In non-dimensional variables, this is equal to $\Phi^{-1/3}\pdv*{\xi}{\tau}$ at leading order in deviatoric strain \citep{webber-2023_linear_formulation} and hence,
    \begin{equation}
        \mathcal{K} = R_1^2 \left.\Phi^{-1/3}\right|_{R=R_1}\dv{R_1}{\tau}.
        \label{eqn:material_flux}
    \end{equation}
    On the inside surface $R=R_1$, requiring no radial stress alongside no normal flow enforces $\epsilon_{rr} = 0$, and therefore
    \begin{equation}
        2(1-\Phi^{1/3}) - \frac{2(R_1-\lambda)}{R_1} = 0 \qq{so} R_1 = \lambda\left.\Phi^{-1/3}\right|_{R=R_1},
        \label{eqn:r1_coreshell}
    \end{equation}
    setting the radius of the core by the local polymer fraction, and making it clear how the inner sphere can freely expand and contract to accommodate changes in the gel coating. Finally, we can solve for $R_2$ using conservation of polymer
    \begin{equation}
        3\int_{R_1}^{R_2}{u^2\Phi\,\mathrm{d}u} = 1-\lambda^3. 
        \label{eqn:conservation_condition}
    \end{equation}
    Together, equation \eqref{eqn:evolution_coreshell} alongside the boundary condition of \eqref{eqn:interfacial_bc}, the material flux in equation \eqref{eqn:material_flux} and the geometrical constraints of \eqref{eqn:r1_coreshell} and \eqref{eqn:conservation_condition} can be used to describe the evolution of the shell of gel as it either deswells or swells. Notice that the stiffness of the sphere at the centre of the device is not taken into account here; we are implicitly considering the case where the inner sphere exerts no mechanical stress on the gel and deforms as required to accommodate the shrinkage or swelling.
    
\begin{figure}
        \centering
        \includegraphics[width=0.9\linewidth]{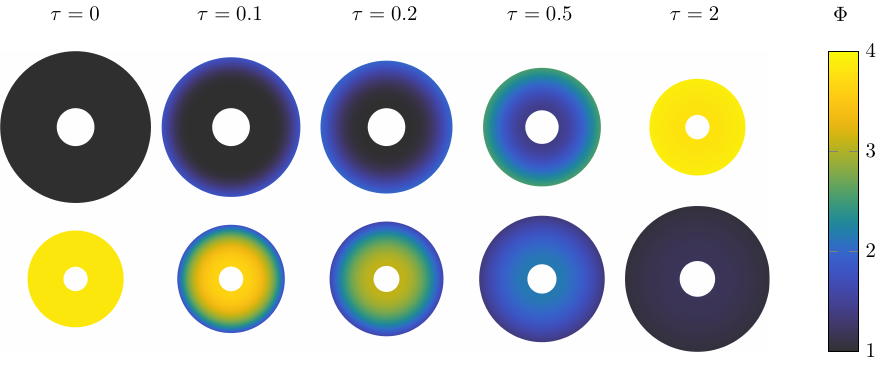}
        \caption{Plots of a responsive gel shell with $\phi_{0\infty} = 4\phi_{00}$, $\lambda = 1/4$ and $\mathcal{M} = 1$ as it deswells from its rest state (top row) or reswells from a `dried' state where $\Phi \equiv \phi_{0\infty}/\phi_{00}$ (bottom row), illustrating how both the polymer fraction $\Phi$ varies throughout the layer but also how the layer thickness and internal and external radii change.}
        \label{fig:coreshell_contour}
    \end{figure}

    To solve this system numerically, introduce the scaled radius
    \begin{equation}
        y = \frac{R-R_1}{R_2-R_1} \qq{with} \pdv{y}{\tau} = -\frac{\dot{R}_1 + (\dot{R}_2-\dot{R}_1)y}{R_2-R_1},
    \end{equation}
    where dots represent derivatives with respect to time, and solve instead on the domain $0\le y \le 1$. Figure \ref{fig:coreshell_contour} illustrates a sample solution to this problem, showing how deswelling and reswelling are not exact reversals of one another and how both the internal and external radius change in time as the gel dries out or rehydrates.

\begin{figure}
        \centering
        \includegraphics[width=0.9\linewidth]{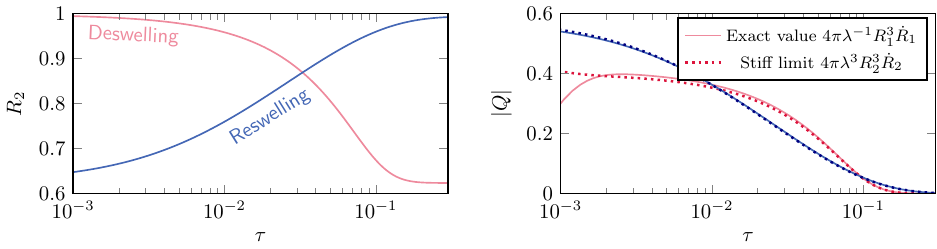}
        \caption{On the left, plots of the outermost scaled radius for a gel layer coating a compressible sphere with $\lambda = 1/5$ and $\mathcal{M}=50$ as it deswells and reswells with partially-dry state $\Phi_0 = 4$. On the right, the magnitude of the fluid fluxes into or out of the same gel, computed exactly using the expression of equation \eqref{eqn:a1_flux} and then approximated in the stiff ($\mathcal{M} \gg 1$) limit using equation \eqref{eqn:a2_flux}.}
        \label{fig:coreshell_plots}
    \end{figure}

    Unlike in the example discussed in section \ref{sec:bz_gel_nd}, the material flux in this case is nonzero and the deswelling or swelling gel layer can therefore act as a source or sink flow, with a strength $Q$, scaled with $\rho_{20}^3/t_{\text{pore}}$, given by
    \begin{equation}
        Q = 4\pi \mathcal{K} = \frac{4\pi}{\lambda}R_1^3\dv{R_1}{\tau}
        \label{eqn:a1_flux}
    \end{equation}
    In the limit of a stiff gel ($\mathcal{M} \gg 1$), equation \eqref{eqn:interfacial_bc} implies that the polymer fraction at the interface is given by $\Phi^{1/3} = 1/R_2$. However, when the gel is stiff, the polymer fraction varies little with radius, and so we can take $\Phi^{1/3} \approx 1/R_2$ throughout the entire thickness of the gel. Since $R_1 = \lambda\Phi^{-1/3}$ in this case, the ratio $R_2/R_1 \approx 1/\lambda$ is preserved throughout the deswelling or swelling process. Thus,
    \begin{equation}
        Q \approx 4 \pi \lambda^3 R_2^3\dv{R_2}{\tau},
        \label{eqn:a2_flux}
    \end{equation}
    allowing the flux from the gel to be quantified entirely in terms of its apparent radius $R_2$ without knowledge of the internal structure. Figure \ref{fig:coreshell_plots} illustrates how the external radius varies in time for a stiff gel, and illustrates how the approximate expression for radial flux in the surrounding fluid \eqref{eqn:a2_flux} is a good fit for the actual flux of equation \eqref{eqn:a1_flux}. After deswelling or reswelling, any steady state solution to equation \eqref{eqn:evolution_coreshell} must have $\Phi$ constant (in order to satisfy boundary conditions). This constant value may not be equal to $\Phi_0$ -- this depends on the value of $\mathcal{M}$ in equation \eqref{eqn:interfacial_bc} -- but in the limit $\mathcal{M} \gg 1$, the system approaches a steady state described by
    \begin{equation}
        \Phi \equiv \Phi_0, \; R_1 = \lambda\Phi_0^{-1/3}, \; R_2 = \Phi_0^{-1/3}.
    \end{equation}

    \section{A Purcell-type two-sphere swimmer}
    \label{sec:purcell_swimmer}
    \begin{figure}
        \centering
        \includegraphics[width=0.9\linewidth]{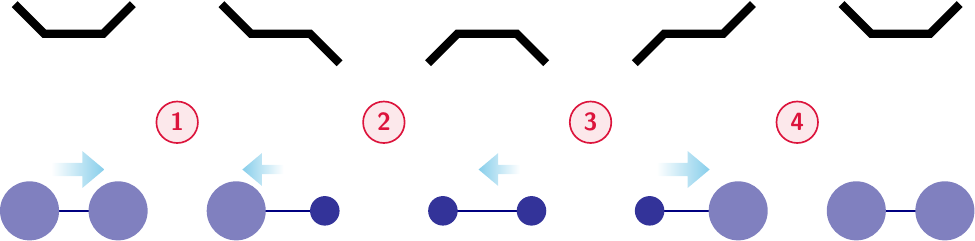}
        \caption{The four stages a Purcell three-link conceptual microswimmer. which drifts to the right through a complete cycle of its actuation \citep{becker-2003_self_swimmer}, compared with the four stages of the two-sphere swimmer, showing how the non-reciprocal nature of the deswelling--reswelling cycle can easily be seen by comparison with Purcell's swimmer. At each stage, the instantaneous velocity of the two-sphere assemblage relative to a fixed frame of reference is shown.}
        \label{fig:swimmers_schematic}
    \end{figure}
    The first non-reciprocal swimmer model introduced by \citet{purcell-1977_life_number} used a non-reciprocal stroke achieved by bending at two hinges, as illustrated in figure \ref{fig:swimmers_schematic}, which leads to a net drift rightwards through the quiescent fluid (for the pattern of actuation illustrated here). Though not the most efficient artificial swimmer to have been designed, running the stroke illustrated in the four steps here backwards leads to a pattern that is reflected horizontally, and so not an exact match of the forwards stroke. Therefore, non-reciprocity is achieved, giving a direction to the resultant motion. In this section, we make the straightforward correspondence between an upward hinge and a swollen gel shell and a downward hinge and a deswollen gel shell to design a non-reciprocal swimmer propelled by the outwards or inwards fluxes of water from the gels.

    Swimmers constructed using two spheres able to change their volume have been considered in the past -- the \textit{pushmepullyou} swimmer \citep{avron-2005_pushmepullyou_swimmer} consists of two spherical bladders connected by a hollow tube, through which an invsicid fluid can be transported, allowing for one sphere to grow at the expense of the other. As a sphere grows, it displaces the background fluid radially, leading to a source-like flow that exerts a viscous drag on the other sphere, pushing it away, whilst the contracting sphere acts like a sink and pulls the other sphere towards it. The combination of these effects leads to locomotion, but physically realising this complex device can be challenging \citep{silverberg-2020_realization_numbers}. A system for transferring interior fluid between the bladders must be designed, and there must be some external actuation to drive the stroke that leads to motion. It is for these reasons that responsive hydrogels provide a more attractive mechanism -- they can be straightforwardly actuated by external stimuli and change their shape without the need for the driven transport of fluid between reservoirs.

    The modelling in sections \ref{sec:bzgel} and \ref{sec:coreshell} give three key insights into self-oscillating gels that we exploit for the modelling in this section, namely:
    \begin{enumerate}
        \item Internal heterogeneity in the chemical concentration field has little effect on the qualitative dynamics, and the only influencing factor is the value of chemical concentration on the water--gel boundary. As such, we take the chemical concentration to be locally homogeneous.
        \item Provided the Damk\"ohler number is sufficiently small, swelling and deswelling will occur in-phase with chemical concentration changes, and sufficiently rapidly that when the chemical concentration next crosses a critical threshold, the swelling or deswelling will be complete, lending the gel a `memoryless' property.
        \item To drive radial fluid flows as a result of deswelling or reswelling, we cannot use a simple uniform hydrogel sphere and must instead consider cases where the material flux $q$ is nonzero. To design a responsive gel swimmer, we employ a sphere constructed from a responsive gel shell coating an impermeable core as modelled in section \ref{sec:coreshell}, but only concern ourselves with the outermost radius $R_2$ since we have shown that this suffices to give information on the nature of the deswelling or reswelling fluxes, at least in the stiff gel $\mathcal{M} \gg 1$ limit where deviatoric stresses are significant.
    \end{enumerate} 
    Somewhat counterintuitively, like the \textit{pushmepullyou} swimmer, deswelling leads to an inwards flux and reswelling to an outwards flux -- even though deswelling involves the expulsion of water, this effect is counteracted by the much greater need to draw water in to occupy space that was previously occupied by gel. A simple four-stage swimming stroke is illustrated below the Purcell swimmer in figure \ref{fig:swimmers_schematic}, where two spheres coated in gel shells of equal swollen and deswollen radii are joined by a rigid rod of fixed length.

    As the right-hand sphere deswells, it draws an inwards flux of fluid that exerts a drag force on the left-hand sphere, pulling it to the right (and dragging the entire system to the right). Then, as the left-hand sphere deswells, the same process exerts a drag on the right-hand sphere, but the entire assembly moves less towards the left as the right-hand sphere has a smaller radius, so feels less of an effect from viscous drag. This step is followed by reswelling of the right-hand sphere, pulling the entire assembly to the left again (since the flow is reversed from step 1), but again with a small effect since the left-hand sphere remains deswollen. Finally, the left-hand sphere reswells, pushing the right-hand sphere rightwards, with a large effect since this first sphere has reswollen. Over the entire course of this actuation, there is a net rightwards drift since these effects (steps 1 and 4) are felt more strongly than the leftwards drift in steps 2 and 3.

    This explanation neglects, however, the drag resisting the motion of the entire system through the fluid, that scales with the sum of the radii of the spheres (since the viscous drag on an individual sphere in a Stokes flow is famously given by $6\pi\mu a \boldsymbol{U}$). If our spheres' radii change by a factor of $0<d<1$ upon deswelling, where $d \approx (\phi_{0\infty}/\phi_{00})^{-1/3}$, and the average scaled radius in a swelling or deswelling cycle is $\bar{d}$ (with $d<\bar{d}<1$), we can use a scaling argument, outlined in table \ref{tab:scaling_purcell}, to show that the effect of the `passive' sphere's changing size is greater than that of the changing drag on the system as a whole, and thus we do see a net motion as described above.

    \begin{table}
        \renewcommand{\arraystretch}{1.5}
        \begin{tabular}{l@{\quad}|@{\quad}c@{\quad}c@{\quad}c@{\quad}c}
            {\color{Crimson}\textbf{Step}} & \circled{1} & \circled{2} & \circled{3} & \circled{4} \\
            Direction & $\rightarrow$ & $\leftarrow$ & $\leftarrow$ & $\rightarrow$ \\
            Relative strength of effect & $1$ & $d$ & $d$ & $1$ \\
            Relative drag on entire system through motion & $1+\bar{d}$ & $d+\bar{d}$ & $d+\bar{d}$ & $1+\bar{d}$ \\
            Relative displacement contribution & $\frac{1}{1+\bar{d}}$ & $\frac{d}{d + \bar{d}}$ & $\frac{d}{d+\bar{d}}$ & $\frac{1}{1+\bar{d}}$
        \end{tabular}
        \caption{An evaluation of the scale of drag forces driving and resisting motion in the two-sphere swimmer of figure \ref{fig:swimmers_schematic}, showing that there is a net rightwards drift for the configuration pictured in figure \ref{fig:swimmers_schematic} even when the changing drag on the system as a whole is taken into consideration. The net displacement will scale with $2/(1+\bar{d}) - 2d/(d+\bar{d}) > 0$.}
        \label{tab:scaling_purcell}
    \end{table}

    \subsection{Motion of the two-sphere swimmer}
    To find the motion of the two-sphere swimmer, we start by considering the motion of two spheres of radii $a_1$ and $a_2$ placed in a quiescent background fluid. The motion of a sphere of radius $a$ placed at the origin into a background steady Stokes flow $\boldsymbol{u_\infty}(\boldsymbol{x})$ can be described exactly by the Fax\'en relation \citep{happel-1983_low_hydrodynamics},
    \begin{equation}
        \boldsymbol{U} = \frac{\boldsymbol{F}}{6\pi\mu_l a} + \boldsymbol{u_\infty}(\boldsymbol{0}) + \frac{a^2}{6}\nabla^2 \boldsymbol{u}_\infty(\boldsymbol{0}),
        \label{eqn:faxen}
    \end{equation}
    where $\boldsymbol{F}$ is the force on the sphere, $\mu_l$ is the fluid viscosity and $\boldsymbol{U}$ is its velocity through the fluid. We model the flow field that results from swelling or drying by placing sources of strength $q_1$ and $q_2$ at the centre of the two spheres, remarking that when deswelling $q_i < 0$ (i.e. there is a sink), like in the swimmer discussed by \citet{avron-2005_pushmepullyou_swimmer}, where an contracting sphere draws fluid inwards radially. It is easy to verify that the velocity field
    \begin{equation}
        \boldsymbol{u} = \frac{q}{4\pi \left|\boldsymbol{x} - \boldsymbol{x_s}\right|^3}\left(\boldsymbol{x} - \boldsymbol{x_s}\right),
    \end{equation}
    both satisfies the Stokes equations and the condition that $q = \int_{\partial \Omega}{\boldsymbol{u}\bcdot\boldsymbol{\hat{n}}}$ for any domain $\Omega$ with outward unit normal $\boldsymbol{\hat{n}}$ surrounding the point $\boldsymbol{x_s}$ -- that is to say, it represents a source of strength $q$ located at this point.

    To deduce the flow field resulting from the movement of both spheres precisely we would need to take into account the changing nature of both the fluxes $q_1(t),\,q_2(t)$ and the radii $a_1(t),\,a_2(t)$, but here we treat them as quasi-steady, assuming that the viscous nature of the surrounding fluid renders time derivatives unimportant. We start by considering the flow due to the motion of the first sphere in a stationary background, with a Stokeslet and point source component, as well as a source of strength $q_1$ located at the origin, which can be added in by linearity,
    \begin{equation}
        \boldsymbol{u} = \frac{3}{4}\boldsymbol{U_1}\left[\frac{a_1}{r} + \frac{a_1^3}{3r^3}\right] + \frac{3}{4 r^2}\left(\boldsymbol{U_1}\bcdot\boldsymbol{x}\right)\boldsymbol{x}\left[\frac{a_1}{r} - \frac{a_1^3}{r^3}\right]+\frac{q_1\boldsymbol{x}}{4\pi r^3} \qq{where} r = \left|\boldsymbol{x}\right|,
        \label{eqn:primitive_sphere_solution}
    \end{equation}
    where $\boldsymbol{U_1}$ is the translation velocity of the first sphere. At leading order, this is the background flow field `felt' by the second sphere, with its centre located at $\boldsymbol{x} = \boldsymbol{\ell}$. The same argument gives an analogous leading order flow field felt by the first sphere at the origin. We could add higher-order corrections, describing how the far-field flow of one sphere is affected by its interactions with the flow induced by the other sphere's motion, using the method of reflections \citep{lamb-1906_hydrodynamics,happel-1983_low_hydrodynamics}, but such corrections would first enter at order-$a_{1,\,2}^4/\ell^4$, where $\ell=\left|\boldsymbol{\ell}\right|$. Thus, the background flow fields close to both spheres are approximated well by
    \begin{subequations}
        \begin{align}
            \boldsymbol{u_\infty^{(1)}} &= \frac{3\boldsymbol{U_2}}{4}\left[\frac{a_2}{r'} + \frac{a_2^3}{3r'^3}\right] + \frac{3}{4 r'^2}\left(\boldsymbol{U_2}\bcdot\boldsymbol{x'}\right)\boldsymbol{x'}\left[\frac{a_2}{r'} - \frac{a_2^3}{r'^3}\right]+\frac{q_2\boldsymbol{x'}}{4\pi r'^3} + \mathcal{O}\left(\frac{a_1^4}{\ell^4}\right)\quad \left(\boldsymbol{x'} = \boldsymbol{x}-\boldsymbol{\ell},\; r' = \left|\boldsymbol{x'}\right|\right) \; \text{and} \\
            \boldsymbol{u_\infty^{(2)}} &= \frac{3\boldsymbol{U_1}}{4}\left[\frac{a_1}{r} + \frac{a_1^3}{3r^3}\right] + \frac{3}{4 r^2}\left(\boldsymbol{U_1}\bcdot\boldsymbol{x}\right)\boldsymbol{x}\left[\frac{a_1}{r} - \frac{a_1^3}{r^3}\right]+\frac{q_1\boldsymbol{x}}{4\pi r^3} + \mathcal{O}\left(\frac{a_2^4}{\ell^4}\right).
        \end{align}
    \end{subequations}
    Since the spheres are connected by a rigid rod, $\boldsymbol{U_1} = \boldsymbol{U_2} = \boldsymbol{U}$. Then, for $\bm{\mathsf{I}}$ the (second-rank) identity tensor,
    \begin{equation}
        \nabla^2 \boldsymbol{u_\infty^{(1)}} = \frac{3a_2}{2r'^3}\, \boldsymbol{U}\bcdot\left(\bm{\mathsf{I}}-\frac{3\boldsymbol{x'x'}}{r'^2}\right) \qq{and} \nabla^2 \boldsymbol{u_\infty^{(2)}} = \frac{3a_1}{2r^3}\, \boldsymbol{U}\bcdot\left(\bm{\mathsf{I}}-\frac{3\boldsymbol{xx}}{r^2}\right).
    \end{equation}
    By symmetry and linearity, both $\boldsymbol{u}$ and $\boldsymbol{F}$ are aligned with $\boldsymbol{\ell}$, and so the Fax\'en relation \eqref{eqn:faxen} gives the swimmer speed aligned with the connecting axis,
    \begin{subequations}
        \begin{align}
            U &= \frac{F_{2 \to 1}}{6\pi \mu_l a_1} + \frac{3U}{4}\frac{a_2}{\ell}\left[1-\frac{2}{3}\left(\frac{a_2}{\ell}\right)^2-\frac{2}{3}\left(\frac{a_1}{\ell}\right)^2\right] - \frac{q_2}{4\pi \ell^2} \qq{and} \\
            U &= \frac{F_{1 \to 2}}{6\pi\mu_l a_2} + \frac{3U}{4}\frac{a_1}{\ell}\left[1-\frac{2}{3}\left(\frac{a_2}{\ell}\right)^2-\frac{2}{3}\left(\frac{a_1}{\ell}\right)^2\right] + \frac{q_1}{4\pi \ell^2},
        \end{align}
    \end{subequations}
    where $F_{1 \to 2}$ is the force on sphere 2 from the connection with sphere 1 and $F_{2 \to 1}$ is the force on sphere 1 from the connection with sphere 2. Owing to the rigidity of the connection, $F_{1 \to 2} = - F_{2 \to 1}$. Hence,
    \begin{equation}
        U = \frac{1}{4\pi \ell^2}\frac{a_2 q_1 - a_1 q_2}{a_1 + a_2}\left[1 - \frac{1}{\ell}\frac{a_1 a_2}{a_1+a_2}\left(\frac{2}{3} - \frac{a_1^2+a_2^2}{\ell^2}\right)\right]^{-1}.
        \label{eqn:swimmer_velocity_q_form}
    \end{equation}
    This expression, accurate up to and including terms of order $a_{1,\,2}^3/\ell^3$, has a leading order contribution from the drag exerted by the swelling or drying fluxes from one sphere on the other, with third-order corrections from the influence of the flow field around one sphere on the other. The entire velocity has an inverse relationship with the sum of the two spheres' radii, representing the drag resisting the motion of the swimmer. Notice that this expression gives the expected flow directions for the swimmer in figure \ref{fig:swimmers_schematic}.
    \begin{figure}
        \centering
        \includegraphics[width=0.9\linewidth]{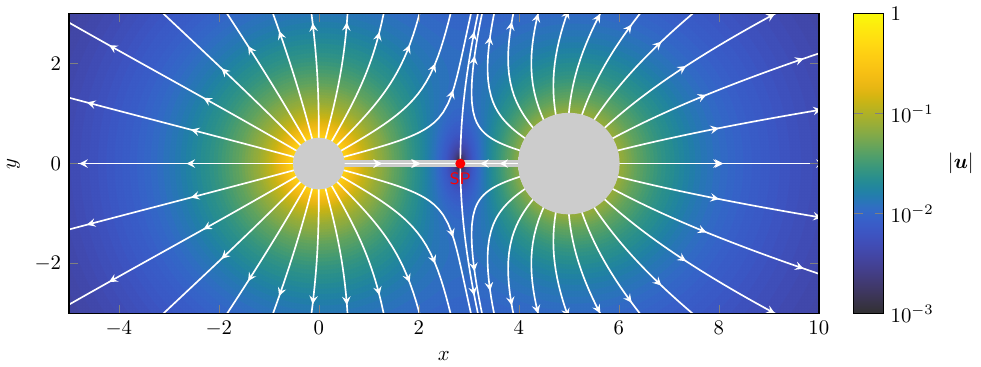}
        \caption{Streamlines around two spheres for a flow with $a_1 = 0.5$, $a_2 = 1$, $\ell=5$, $q_1=1.5$ and $q_2=1$ (normalised units, with all lengths scaled by $a_0$ and times scaled arbitrarily). The stagnation point is marked along the $x$ axis. In this case, $U = 2.2 \times 10^{-3}$, with a net motion to the right.}
        \label{fig:streamlines}%
    \end{figure}

    The approximate flow field around the swimmer can be computed by taking a linear superposition of the flow due to each sphere of the form \eqref{eqn:primitive_sphere_solution}, substituting for the velocity of the total assemblage by using the expression of equation \eqref{eqn:swimmer_velocity_q_form},
    \begin{equation}
        \boldsymbol{u} \approx \frac{3(a_2 q_1 - a_1 q_2)\boldsymbol{\hat{x}}}{16\pi \ell^2}\bcdot\frac{\left(\frac{a_1}{r} + \frac{a_2}{r'} + \frac{a_1^3}{3r^3} + \frac{a_2^3}{3r'^3}\right)\bm{\mathsf{I}} + \left(\frac{a_1}{r^3} + \frac{a_2}{r'^3} - \frac{a_1^3}{r^5} - \frac{a_2^5}{r'^3}\right)\boldsymbol{x}\boldsymbol{x}}{a_1 + a_2 - \frac{a_1 a_2}{\ell}\left(\frac{2}{3} - \frac{a_1^2+a_2^2}{\ell^2}\right)}+\frac{q_1\boldsymbol{x}}{4\pi r^3}+\frac{q_2\boldsymbol{x'}}{4\pi r'^3}
    \end{equation}
    This solution approximately satisfies the no-penetration boundary conditions on $r=a_1$ and $r'=a_2$, and a sample such flow field is plotted in figure \ref{fig:streamlines}. The form of this solution also provides a post-hoc justification for our neglect of the effect of the rod that joins the two spheres on the dynamics of the swimmer. If $\boldsymbol{x} \parallel \boldsymbol{\hat{x}}$, it is clear that $\boldsymbol{u} \parallel \boldsymbol{\hat{x}}$, and no flow crosses the position of the thin rod. The disturbance to the background flow decays like $1/r$ at leading order, with a magnitude proportional to the fluxes and inversely proportional to the square of the separation between spheres, $\ell$, with a potential stagnation point along the line joining the centre of the two spheres. Finally, equation \eqref{eqn:a2_flux}, rewritten in dimensional terms, shows that
    \begin{equation}
        q_i = \frac{4 \pi \lambda^3}{(\rho_{20})_i}a_i^3\dv{a_i}{t},
    \end{equation}
    where $(\rho_{20})_i$ is the rest state outer radius of sphere $i$ and $\lambda_i$ is the gel layer thickness parameter. Hence, the speed of the overall swimmer can be written in terms of the changing radii,
    \begin{equation}
        U = \frac{1}{3 \ell^2}\frac{a_1 a_2}{a_1+a_2}\left[1 - \frac{1}{\ell}\frac{a_1 a_2}{a_1+a_2}\left(\frac{2}{3} - \frac{a_1^2+a_2^2}{\ell^2}\right)\right]^{-1}\dv{t}\left(\frac{\lambda_1^3 a_1^3}{(\rho_{20})_1}-\frac{\lambda_2^3 a_2^3}{(\rho_{20})_2}\right).
    \end{equation}

    \subsection{Drift motion}
\begin{figure}
        \centering
        \includegraphics[width=0.9\linewidth]{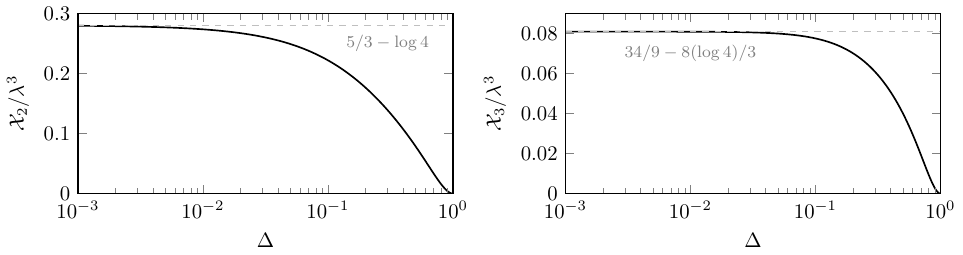}
        \caption{Plots of the second-order ($\mathcal{X}_2$) and third-order ($\mathcal{X}_3$) drift magnitudes as shown in equations \eqref{eqn:x_2} and \eqref{eqn:x_3} against degree of deswelling $\Delta = d/a_0$, showing larger drifts for greater deswelling. The asymptotic values as $\Delta \to 0$ are shown with the dashed lines.}
        \label{fig:drifts}%
    \end{figure}

    Assume that the natural fully-swollen radius of both spheres is $a_0 = (\rho_{20})_1 = (\rho_{20})_2$ such that the poroelastic timescale $t_{\text{pore}} = \mu_l a_0^2/k\Pi_0$. Then introduce the non-dimensional quantities
    \begin{equation}
        \tau = \frac{t}{t_{\text{pore}}}, \quad \mathcal{U} = \frac{t_{\text{pore}}}{a_0}U \qq{and} A_{1,\,2} = \frac{a_{1,\,2}}{a_0},
    \end{equation}
    and let $\delta = a_0/\ell$. Then,
    \begin{align}
        \mathcal{U} &= \frac{\delta^2}{3}\frac{A_1 A_2}{A_1+A_2}\left[1-\frac{2\delta}{3}\frac{A_1 A_2}{A_1+A_2} + \delta^3 \frac{A_1A_2\left(A_1^2+A_2^2\right)}{A_1+A_2}\right]^{-1}\dv{\tau}\left[\lambda_1^3A_1^3-\lambda_2^3A_2^3\right] \notag \\
        &= \frac{\delta^2}{3}\frac{A_1 A_2}{A_1+A_2}\dv{\tau}\left[\lambda_1^3A_1^3-\lambda_2^3A_2^3\right] + \frac{2\delta^3}{9}\frac{A_1^2 A_2^2}{(A_1+A_2)^2}\dv{\tau}\left[\lambda_1^3A_1^3-\lambda_2^3A_2^3\right] + O(\delta^4),
        \label{eqn:velocity}
    \end{align}
    where we neglect contributions of order $\delta^4$ and above, since these are the same size as the `reflection' terms we neglected when calculating the flow field, and so cannot be determined accurately in this model. Thus, the instantaneous drift velocity that we consider only has contributions from the flow field of each sphere individually and the fluxes from the sources, with no effects from sphere--sphere flow interactions. Over one stroke $0 \le \tau \le T$, there is a dimensionless drift $\mathcal{X}$ given by
    \begin{equation}
        \mathcal{X} = \int_0^T{\mathcal{U}\,\mathrm{d}\tau} \qq{with} \mathcal{X} = \delta^2 \mathcal{X}_2 + \delta^3 \mathcal{X}_3 + \dots.
    \end{equation}
    To compute the leading-order drift $\mathcal{X}_2$ for the symmetric Purcell--type swimmer carrying out the stroke illustrated in figure \ref{fig:swimmers_schematic} with $\lambda_1=\lambda_2=\lambda$, let $\Delta = d/a_0$ and thus, integrating over the four strokes of the cycle,
    \begin{align}
        \mathcal{X}_2 &= -\lambda^3\int\limits_{\smlcircled{1}}{\frac{A_2^3}{1+A_2}\dv{A_2}{\tau}\,\mathrm{d}\tau} + \lambda^3\int\limits_{\smlcircled{2}}{\frac{\Delta A_1^3}{\Delta+A_1}\dv{A_1}{\tau}\,\mathrm{d}\tau} - \lambda^3\int\limits_{\smlcircled{3}}{\frac{\Delta A_2^3}{\Delta+A_2}\dv{A_2}{\tau}\,\mathrm{d}\tau} + \lambda^3\int\limits_{\smlcircled{4}}{\frac{A_1^3}{1+A_1}\dv{A_1}{\tau}\,\mathrm{d}\tau} \notag \\
        &=2\lambda^3\int_\Delta^1{\frac{u^3}{1+u} - \frac{\Delta u^3}{\Delta + u}\,\mathrm{d}u} = 2\lambda^3\left[\log{\left(\frac{1+\Delta}{2}\right)}+\Delta^4\log{\left(\frac{1+\Delta}{2\Delta}\right)} + \frac{1}{6}(1-\Delta)^2\left(5\Delta^2+2\Delta+5\right)\right].
        \label{eqn:x_2}
    \end{align}
    An analogous calculation gives the third-order correction
    \begin{equation}
        \mathcal{X}_3 = \frac{16 \lambda^3}{3}\left[\log{\left(\frac{1+\Delta}{2}\right)}+\Delta^5\log{\left(\frac{1+\Delta}{2\Delta}\right)}\right]+\frac{2\lambda^3}{9}\frac{(1-\Delta)^2}{1+\Delta}\left(17\Delta^4 + 27\Delta^3+23\Delta^2+27\Delta+17\right).
        \label{eqn:x_3}
    \end{equation}
    The drift is plotted against $\Delta$ in figure \ref{fig:drifts}, showing how there is, as expected, no drift when $\Delta \to 1$, corresponding to no deswelling, and a maximal drift as $\Delta \to 0$. Notice also that $\mathcal{X} \propto \lambda^3$, and so the drift is greater for thinner layers of gel, where deswelling relies on expulsion of water and an expansion of the inner sphere, relative to thicker spherical shells $\lambda \to 0$ where deswelling expels less fluid. In the $\lambda \to 0$ limit, the case of a sphere with no net fluxes is retrieved, and the swimmer does not move.

    \section{An in-phase nonreciprocal swimmer}
    \label{sec:in_phase}
    \begin{figure}
        \centering
        \includegraphics[width=0.9\linewidth]{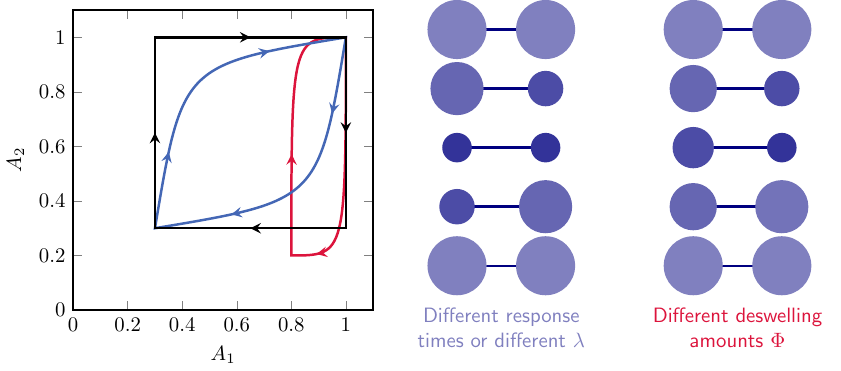}
        \caption{Illustrations of the path of a single stroke in $A_1-A_2$ phase space showing the non-reciprocal nature through the nonzero area swept out. The black curve illustrates the path of the `Purcell' swimmer, where deswelling and reswelling of each sphere occurs out of phase with the other. Two synchronous alternatives are presented: gels which respond at different rates, either due to material properties or different thicknesses $\lambda$, and gels that deswell to a greater or lesser degree when the local conditions are changed.}%
        \label{fig:phase}%
    \end{figure}%
    The swimmer outlined in the previous section is of limited utility if the two shape-changing spheres are formed from identical chemically-responsive hydrogels: on the scale of a microswimmer, we expect the chemical concentration field to be approximately uniform, and so mechanisms for out-of-phase actuation where one gel deswells but the other does not are likely challenging to replicate in experiments. To achieve a net drift, a nonzero area must be traced out in $A_1-A_2$ phase space over the course of one stroke \citep{golestanian-2008_analytic_number}. This is straightforwardly achieved by the Purcell--like swimmer of the previous section, where the trajectory is square (figure \ref{fig:phase}). It is therefore important that there is an asymmetry between deswelling and reswelling processes, and we illustrate two ways in which this can occur in figure \ref{fig:phase}. We focus on the first of these approaches in the analysis that follows, where the rest radii for the two gel spheres are the same in the swollen and deswollen limits, and only the transient dynamics change.

    To overcome this shortcoming, we consider a swimmer where both geometrically identical gels respond at the same time, albeit at different rates. This rate difference preserves the irreversibility necessary for a nonreciprocal motion, and can be summarised by
    \begin{equation}
        \centering
        \begin{tikzpicture}[baseline=(current  bounding  box.center)]
            \node at (-2, 0) {$(A_1,\,A_2)$};
            \node at (-0.9, 0) {$=$};
            \node (s1) at (0, 0) {$(1,\,1)$};
            \node (s2) at (3, 0) {$(\Delta,\,\Delta)$};
            \node (s3) at (6, 0) {$(1,\,1)$.};

            \draw[color=Crimson, thick, -{Stealth}] (-0.2, -0.25) .. controls (1, -0.75) and (2, -0.75) .. (2.8, -0.25);
            \node[color=Crimson] at (1.3, -1) {\textsf{\textit{slow}}};
            \draw[color=Crimson, very thick, -{Stealth}] (0.2, 0.25) .. controls (1, 0.75) and (2, 0.75) .. (3.2, 0.25);
            \node[color=Crimson] at (1.5, 1) {\textsf{\textit{fast}}};

            \draw[color=Navy!50!SkyBlue, thick, -{Stealth}] (2.85, -0.25) .. controls (4, -0.75) and (5, -0.75) .. (5.8, -0.25);
            \node[color=Navy!50!SkyBlue] at (4.325, -1) {\textsf{\textit{slow}}};
            \draw[color=Navy!50!SkyBlue, very thick, -{Stealth}] (3.25, 0.25) .. controls (4, 0.75) and (5, 0.75) .. (6.2, 0.25);
            \node[color=Navy!50!SkyBlue] at (4.525, 1) {\textsf{\textit{fast}}};
        \end{tikzpicture}
    \end{equation}

    The response time of a hydrogel to a change in stimulus, $t_{\text{pore}}$, is limited by viscous resistance hampering the fluid flow through the porous structure of the gel. The permeability of these gels can be as low as $10^{-15}\,\mathrm{m}^2$, setting an upper limit on response rates, but this can be modified by introducing micropores into the gel through which water can be driven out or drawn in much more rapidly \citep{spratte-2022_thermoresponsive_microchannels}. This allows us to engineer different response times for two gels even if they are constructed from the same responsive gel material and otherwise would deswell to the same extent. 
    
    Let the characteristic response time for the first gel be $t_1 = t_{\text{pore}}$ and, for the second gel, $t_2$ with $t_1/t_2 = \mathcal{T} \gg 1$. This allows us to introduce a slow timescale $\tau = t/t_1$ and a fast timescale $\tau_f = t/t_2$ such that $A_1 = A_1(\tau)$ and $A_2 = A_2(\tau_f)$. Then, equation \eqref{eqn:velocity} becomes
    \begin{equation}
        \frac{\mathcal{U}}{\delta^2 \lambda^3} = \left[\frac{A_1 A_2}{A_1+A_2} + \frac{2\delta}{3}\frac{A_1^2A_2^2}{(A_1+A_2)^2}\right]\left(A_1^2 \dv{A_1}{\tau} - \mathcal{T}A_2^2 \dv{A_2}{\tau_f}\right) \approx -\mathcal{T}\left[\frac{A_1 A_2^3}{A_1+A_2} + \frac{2\delta}{3}\frac{A_1^2A_2^4}{(A_1+A_2)^2}\right]\dv{A_2}{\tau_f} + \dots,
    \end{equation}
    since $\mathcal{U}$ is scaled with the slow (standard poroelastic) timescale. Thus, during the deswelling part of the stroke, $A_1 \approx 1$ and during the reswelling part $A_1 \approx \Delta$, so that
    \begin{equation}
        \frac{\mathcal{X}}{\mathcal{T}\delta^2\lambda^3} \approx -\int^\Delta_{1}{\left[\frac{u^3}{1+u} + \frac{2 \delta}{3}\frac{u^4}{(1+u)^2}\right]\,\mathrm{d}u} - \int_\Delta^{1}{\left[\frac{u^3}{\Delta+u} + \frac{2 \delta \Delta}{3}\frac{u^4}{(\Delta+u)^2}\right]\,\mathrm{d}u},
        \label{eqn:drift_synchronous}
    \end{equation}
    which evaluates to the same drift as that calculated in equations \eqref{eqn:x_2} and \eqref{eqn:x_3}, albeit with a corrective prefactor of $\mathcal{T}$, showing how this swimmer, at leading order, is equivalent to that of the Purcell-like out-of-phase swimmer. One stroke achieves a greater drift if the separation between timescales is greater, as would be expected, as we trace out a larger area in the phase space of figure \ref{fig:phase} in this case.

\begin{figure}
        \centering
        \includegraphics[width=0.9\linewidth]{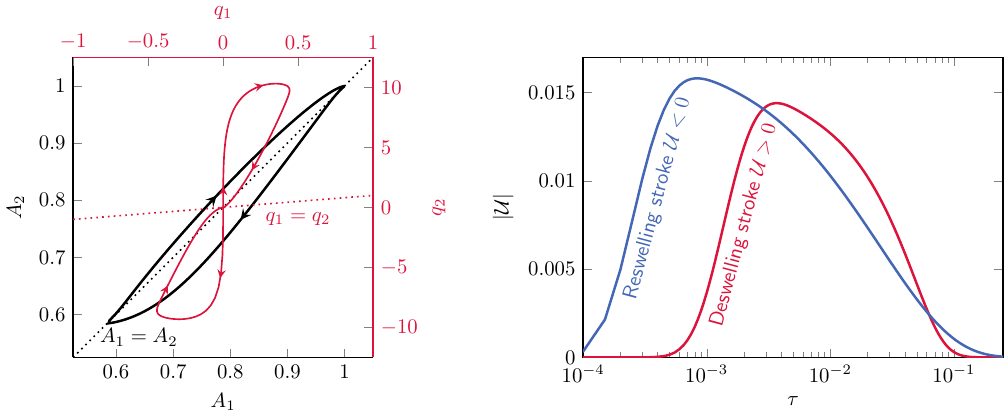}
        \caption{Plots of an asymmetric two-sphere swimmer where $\lambda$ differs between the two spheres, with $\lambda_1 = 1/5$ and $\lambda_2 = 1/2$ and $\delta = 1/5$, as well as $\phi_{0\infty}=5\phi_{00}$. On the left, the stroke paths in phase space for such a swimmer, showing a nonzero area traced out and the variation of fluxes $q_i$ through both deswelling and reswelling. On the right, for the same swimmer, the instantaneous velocities are plotted to show that the backward stroke (when reswelling) is not an exact inverse of the forward stroke, and there is in fact a net rightwards drift -- though not immediately apparent on the logarithmic axis, the area under the red curve is greater than that under the blue curve.}
        \label{fig:real_swimmer}%
    \end{figure}

    Alternatively, we can link together two spheres with different values of $\lambda$ and thus different thicknesses of gel coating. If both spheres have the same initial radius, they will deswell to the same radii in the $\mathcal{M} \gg 1$ limit, but the transient evolution will be different in each case. Using the formulation of section \ref{sec:coreshell}, we can calculate the evolution of the radii and the fluxes and show that a closed path of nonzero area in $A_1-A_2$ phase space is traced out, with the resultant forwards stroke velocity not being an exact opposite of the backwards stroke velocity, resulting in a net drift, as illustrated in figure \ref{fig:real_swimmer}.

    \subsection{Approximate (fitted) gel radii}
    \begin{figure}
        \centering
        \includegraphics[width=0.9\linewidth]{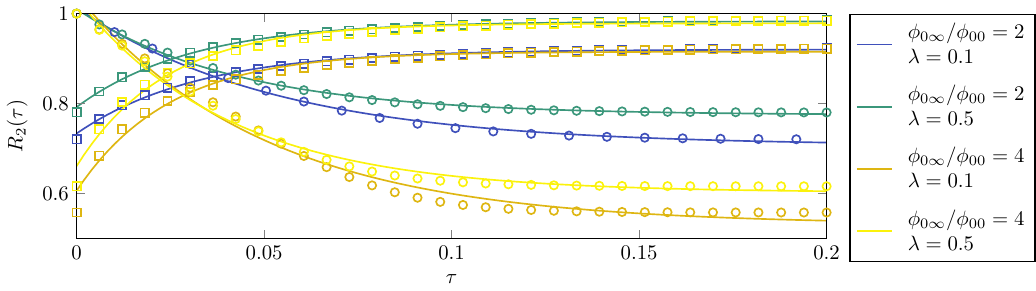}
        \caption{Plots of the evolution of the outermost radius in time when swelling and deswelling for gel-coated spheres with different equilibrium polymer fractions and gel thicknesses $\lambda$, showing the applicability of an exponential fit $R_2 = R_2^{\text{end}} + \left(R_2^{\text{start}}-R_2^{\text{end}}\right)e^{-\kappa\tau}$. Here, $\mathcal{M}=50$, circles represent deswelling and squares reswelling, and the exponential fits (found using Matlab's \texttt{lsqnonlin}) are plotted as solid lines.}
        \label{fig:exp_fit}
    \end{figure}
    For the sake of straightforward analytic treatment, we restrict our attention to the swimmer with different response times but the same value of $\Delta$ (and $\lambda$) for each sphere, though the analysis that follows could easily be repeated for spheres which deswell to different extents, or which have different gel thicknesses. In this case, we can vary the separation of response timescales, $\mathcal{T}$ to tune the degree of asymmetry in the swimmer, expecting faster drifts for larger separations $\mathcal{T}$. Starting from the earliest theoretical studies of gel dynamics, such as the dominant-mode expansion of \citet{tanaka-1979_kinetics_gels}, it is well-known that the radius of a swelling or drying hydrogel sphere can be approximated, at leading order, by an exponential that decays on the timescale of poroelastic reconfiguration. The same applies in the case of our gel shell, as evidenced by the plots of figure \ref{fig:exp_fit}, showing that both swelling and drying follow this exponential decay model,
    \begin{equation}
        R_2 = R_2^{\text{end}} + \left(R_2^{\text{start}}-R_2^{\text{end}}\right)e^{-\kappa\tau},
    \end{equation}
    where $\kappa$ is a function of $\lambda$, $\phi_{0\infty}$ and whether the gel is deswelling or reswelling. However, for simplicity, we take $\kappa \equiv 1$ in the analysis that follows, remarking that simply rescaling time variables can achieve this same result, and that the range of decay rates $\kappa$ is not large.

\begin{figure}
        \centering
        \includegraphics[width=0.9\linewidth]{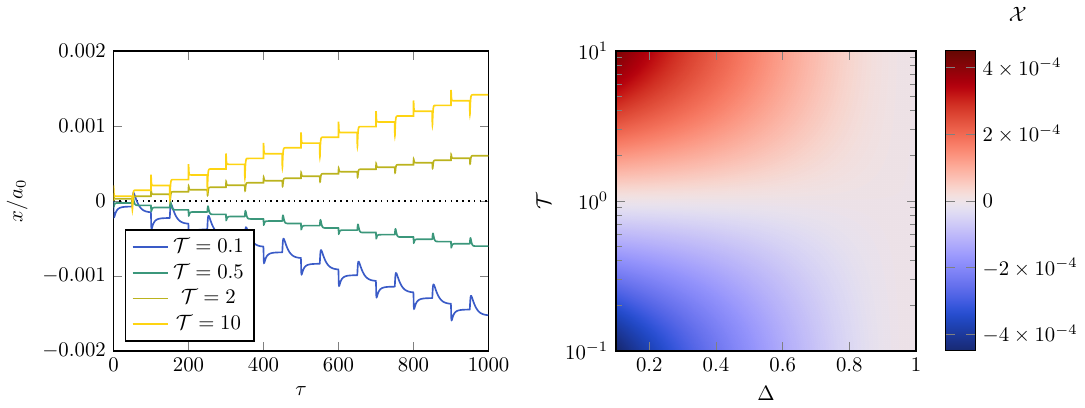}
        \caption{On the left, plots of the location of a swimmer with $\Delta = \lambda = 0.5$ for various values of $\mathcal{T}$ after repeated strokes of length $\tau=100$, with the first half of the stroke corresponding to deswelling and the second half to reswelling. On the right, a plot of the drift magnitudes as a function of the separation of timescales $\mathcal{T}$ and deswollen radius $\Delta$. For both plots, $\delta^2 = 1/50$.}
        \label{fig:paths_swimmer}%
    \end{figure}

    The precise radial structure of the polymer fraction field can be derived from the advection--diffusion equation of any poroelastic model \citep{punter-2020_compression_model}, but we only need to know the radius of the sphere in order to deduce the magnitude of the drag forces and the size of the deswelling or reswelling fluxes. Then, on the deswelling part of the stroke,
    \begin{equation}
        A_1 = \Delta + (1-\Delta)\operatorname{exp}{\left(-\tau\right)} \qq{and} A_2 = \Delta + (1-\Delta)\operatorname{exp}{\left(-\tau_f\right)},
        \label{eqn:fitted_deswell}
    \end{equation}
    whilst on the reswelling part,
    \begin{equation}
        A_1 = 1 + (\Delta-1)\operatorname{exp}{\left(-\tau\right)} \qq{and} A_2 = 1 + (\Delta-1)\operatorname{exp}{\left(-\tau_f\right)}.
        \label{eqn:fitted_reswell}
    \end{equation}
    Hence, the second-order (in $\delta$) instantaneous velocity can be computed
    \begin{subequations}
        \begin{align}
            \mathcal{U}_{\text{deswell}} &= \frac{\delta^2 \lambda^3 (1-\Delta)f_1(\tau_f)f_1(\tau)}{2\Delta + (1-\Delta)\left(e^{-\tau_f}+e^{-\tau}\right)}\left[\mathcal{T}f_1(\tau_f)^2e^{-\tau_f} - f_1(\tau)^2e^{-\tau}\right] \qq{with} f_1(x) = \Delta + (1-\Delta)e^{-x}, \label{eqn:deswell_velocity}\\
            \mathcal{U}_{\text{reswell}} &= \frac{\delta^2 \lambda^3 (\Delta-1)f_2(\tau_f)f_2(\tau)}{2 + (\Delta-1)\left(e^{-\tau_f}+e^{-\tau}\right)}\left[\mathcal{T}f_2(\tau_f)^2e^{-\tau_f} - f_2(\tau)^2e^{-\tau}\right] \qq{with} f_2(x) = 1 + (\Delta-1)e^{-x}. \label{eqn:reswell_velocity}
        \end{align}
    \end{subequations}
    These expressions for $\mathcal{U}$ allow us to deduce how the position of a swimmer changes throughout its stroke, with both forwards and backwards motion as shown in figure \ref{fig:paths_swimmer}. Notice that the drift motion reverses when $\mathcal{T} \to 1/\mathcal{T}$ as the timescale separation occurs in the other direction (i.e. the first sphere deswells more rapidly). It is clear that the magnitude of the drift will increase for larger amounts of deswelling (smaller $\Delta$), and figure \ref{fig:paths_swimmer} shows how a greater separation of timescales also leads to a bigger drift.

    \section{Surfing and bobbing}
    \label{sec:surf_bob}
    So far, we have considered swimmers in a spatially-uniform chemical field that locomote as chemical signals are repeatedly switched on and off, and they feel the same signals no matter where they are located in space. Further behaviours are apparent if we allow the chemical concentration field to vary in space, and then the motion of the swimmers themselves can feed back into the forcing signal that drives a swimming stroke. We can imagine scenarios where oscillating chemical fields drive swimmers into regions where these oscillations no longer occur, thus trapping them, or where initially stationary swimmers are met by a travelling front of chemical concentration, leading to motion. Such chemical waves can be generated by reaction-diffusion systems such as the BZ reaction, as detailed for example in a single spatial dimension in appendix \ref{app:bz_waves}. In the analysis that follows, we will assume for simplicity that neither the deswelling and reswelling behaviour of the swimmer, nor its motion through the background fluid, will influence the processes that generate these stimulus waves.

    For simplicity's sake, consider planar chemical waves travelling in the $z$ direction, described by a series of fronts travelling at speed $c_0$, described by
    \begin{equation}
        z - c_0 \tau = f n \qq{with} n \in \mathbb{Z},
    \end{equation}
    where $f/c_0$ represents the (dimensionless) duration of the pulse, and $f$ is a (half) wavelength. The chemical concentration alternates between values above and below the critical value $Y_c$ as each front passes, leading to periodic deswelling and reswelling. When the chemical front travelling at speed $c_0$ coincides with the position of a stationary, fully swollen, swimmer at time $\tau_0$ and position $z_0$, the an approximation to the instantaneous deswelling velocity $\mathcal{U}_0^I$ can be computed using equation \eqref{eqn:deswell_velocity}, with
    \begin{equation}
        \mathcal{U}_0^I = \frac{\delta^2 \lambda^3 (\mathcal{T}-1)}{2} (1-\Delta) \qq{so} z \approx z_0 + \frac{\delta^2 \lambda^3 (\mathcal{T}-1)}{2} (1-\Delta)(\tau-\tau_0)
        \label{eqn:first_criterion}
    \end{equation}
    for times soon after $\tau_0$. The subsequent behaviour depends on the nature of the chemical concentration field at this new position -- either the swimmer continues to deswell, or it outpaces the chemical front and instead reswells back towards its equilibrium state. We refer to these two behaviours as `bobbing' and `surfing', respectively.
    
    \begin{figure}
        \centering
        \includegraphics[width=0.9\linewidth]{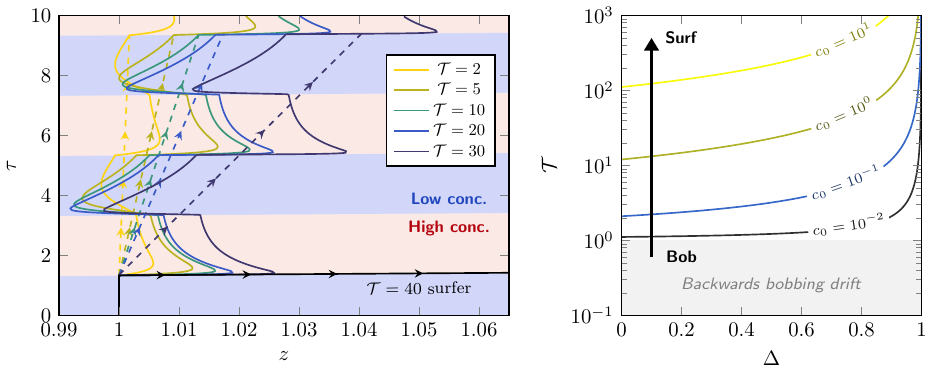}
        \caption{On the left, plots of position of a swimmer with $\Delta=0.5$ that starts at $z=1$ as chemical waves with speed $c_0 = 1$ pass. For low values of timescale separation $\mathcal{T}$, the initial deswelling at $\tau = 1$ is slow enough that the swimmer will settle into a `bobbing' pattern, with a net drift over multiple cycles as illustrated by the dashed lines. For large enough $\mathcal{T}$, the swimmer can `surf' along the chemical wave at speed $c_0$. On the right, the boundary between regions where swimmers `bob' and `surf' in $\Delta-\mathcal{T}$ space is illustrated for different chemical wave speeds. For these plots, we have taken $\delta = 0.5$, $\lambda = 0.9$ and $c_0 = 0.75$.}
        \label{fig:bob_v_surf}%
    \end{figure}

    \subsection{Bobbing}
    Provided that the chemical front travels faster than the instantaneous deswelling--driven velocity of the swimmer ($c_0 > \mathcal{U}_0^I$), the swimmer remains in the region of $z-\tau$ space where it deswells, and its motion is described by equation \eqref{eqn:deswell_velocity}. Eventually, another chemical front catches up with the swimmer and the background chemical field returns to below the critical deswelling threshold, so the spheres reswell and the swimmer travels with velocity given by equation \eqref{eqn:reswell_velocity}. Together, these two phases of behaviour describe one stroke of swimmer motion as discussed in the previous sections, and there is a net drift of magnitude $\mathcal{X}$ over this entire cycle of duration $2f/c_0$. Since this drift must be order-$\delta^2$, the leading-order drift velocity can be approximated by
    \begin{equation}
        \mathcal{U}_d = \frac{c_0\mathcal{X}}{2 f} \approx \frac{\delta^2 c_0 \mathcal{X}_2}{2 f} \approx \frac{\delta^2  \lambda^3 c_0 \mathcal{T}}{f}\left[\log{\left(\frac{1+\Delta}{2}\right)}+\Delta^4\log{\left(\frac{1+\Delta}{2\Delta}\right)} + \frac{1}{6}(1-\Delta)^2\left(5\Delta^2+2\Delta+5\right)\right],
    \end{equation}
    as $\mathcal{T}\to\infty$, using the analysis of equation \eqref{eqn:drift_synchronous}.
    
    This motion can be viewed as analogous to a floating tracer on the surface of a water bath undergoing oscillatory motion, with a net drift in the direction of wave travel -- even though the underlying mechanisms differ significantly, the second-order drift that results is qualitatively similar to the famous Stokes drift \citep{stokes-1847_theory_waves} that results when surface waves propagate in water. This very same oscillatory behaviour is already exploited in experimental devices that can transport droplets of material along a surface \citep{murase-2008_design_gel}, where the material `bobs' as peristaltic contractile waves pass. Figure \ref{fig:bob_v_surf} illustrates some example bobbing swimmers, that drift in the direction of chemical wave travel as chemical waves pass.

    \subsection{Surfing}
    Notice that, as $\mathcal{T}$ increases, the drift velocities rapidly increase: the swimmer `catches' the chemical wave and immediately deswells, travelling ahead of the wave and ceasing its deswelling, only to be caught by the wave again and for the process to repeat. In this case, the motion cannot be described purely as a `bobbing', since there are some periods of time where the swimmer travels along with the chemical wave, an entirely different net transport mechanism.

    A short time after the chemical front passes, its position can be described by $z_0 + c_0(\tau-\tau_0)$, and therefore if $c_0 < \mathcal{U}_0^I$, the swimmer finds itself ahead of the front immediately after it begins to deswell. In these surroundings, the chemical concentration is below the critical threshold for deswelling, so the deswelling will stop and the spheres may actually take on more fluid to return to their equilibrium scaled radii $A_1=A_2=1$. Therefore, there will be no more motion of the swimmer, and it will remain fixed in place until the front catches up, at which point the process will repeat. This mechanism allows swimmers to `surf' along the chemical front at speed $c_0$, potentially allowing for much faster travel than by pure swimming alone. This occurs for larger values of $\mathcal{T}$ in figure \ref{fig:bob_v_surf}, but cannot happen indefinitely: in order to propel itself forwards when caught by the chemical wave, the second sphere necessarily loses some water in deswelling which it does not regain before the chemical front catches up again. This means that the subsequent outwards flux is reduced, and this process repeats until the swimmer can no longer propel itself forwards at the requisite speed for surfing, as seen in the $\mathcal{T}=20$ and $\mathcal{T}=30$ plots, where the subsequent behaviour is more akin to bobbing. In other words, the stable propulsion method for all such swimmers is bobbing, with surfing a much faster, yet inherently unstable, approach available to some.

    \subsubsection{Transition from surfing to bobbing}
\begin{figure}
        \centering
        \includegraphics[width=0.9\linewidth]{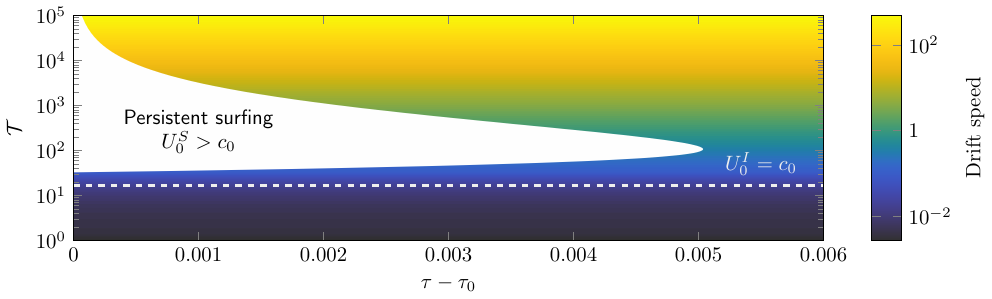}
        \caption{A phase diagram for a swimmer with $\delta = \Delta = 1/2$, $\lambda=0.9$ and $c_0 = 0.75$ and $f=2$ showing the possible behaviours exhibited as time progresses. There can be persistent surfing if $U_0^S > c_0$, transient surfing at early times when $U_0^I > c_0$, and otherwise bobbing at a drift speed that is shown on the plot.}
        \label{fig:surfing_breakdown}%
    \end{figure}

    When in the surfing regime, spheres only deswell for instantaneously small time periods, and reswell when they exceed the speed of the propagating chemical front. In the analysis that follows, we consider only the changing shape of the faster-reacting second sphere (the $\mathcal{T} \gg 1$ limit), and seek to understand the criteria for a surfing sphere to exist that does not lose enough fluid during its deswelling to return to bobbing behaviour. If the radius of the sphere at some instant when it catches the chemical wave is $A_2^0$ , the subsequent evolution can be described by
    \begin{equation}
        A_2^{(d)} = \Delta + (A_2^0-\Delta)\exp[-(\tau_f-\mathcal{T}\tau_0)] \qq{or} A_2^{(r)} = 1 + (A_2^0-1)\exp[-(\tau_f-\mathcal{T}\tau_0)],
    \end{equation}
    dependent on whether we are deswelling or reswelling. When deswelling, $\dv*{A_2}{\tau_f} \approx -(A_2^0-\Delta)$, whilst when reswelling $\dv*{A_2}{\tau_f} \approx (1 - A_2^0)$, a much smaller effect since $A_2$ is close to $1$. It is natural to assume that the sphere spends half of its time deswelling and half of it reswelling, so that the actual rate of deswelling can be approximated by the average of these two regimes
    \begin{equation}
        \dv{A_2}{\tau_f} \approx -\left(A_2 - \frac{1+\Delta}{2}\right)\exp[-(\tau_f-\mathcal{T}\tau_0)] \qq{so} A_2 \approx \frac{1+\Delta}{2} + \frac{1-\Delta}{2e}\exp\left(e^{-(\tau_f-\mathcal{T}\tau_0)}\right),
    \end{equation}
    giving a much slower, but still appreciable, rate of deswelling whilst surfing. Assuming, then, that $A_1$ remains close to $1$ throughout, the swimmer velocity is approximately
    \begin{equation}
        \mathcal{U}_0^S \approx \frac{\lambda^3 \delta^2 \mathcal{T} (1-\Delta)}{8}\frac{\left[1+\Delta + (1-\Delta)\exp\left(e^{-\mathcal{T}(\tau-\tau_0)}-1\right)\right]^3}{3 + \Delta + (1-\Delta)\exp\left(e^{-\mathcal{T}(\tau-\tau_0)}-1\right)}\exp\left(e^{-\mathcal{T}(\tau-\tau_0)}-\mathcal{T}(\tau-\tau_0) - 1\right),
        \label{eqn:second_criterion}
    \end{equation}
    with $\tau_0$ the time that surfing begins. Surfing continues to happen at such times $\tau$ where this approximate velocity is greater than $c_0$ and ceases when this is not the case. As $\tau \to \infty$, it is clear that $\mathcal{U}_0^S \to 0$, as would be expected and the sphere has deswollen to an equilibrium value -- it is of course the case that a bobbing regime will always be reached at the very latest times. However, for small $\tau-\tau_0$,
    \begin{equation}
        \mathcal{U}_0^S \approx \frac{\delta^2 \lambda^3 \mathcal{T}}{4}(1-\Delta),
    \end{equation}
    a much more stringent condition for surfing than that in condition \eqref{eqn:first_criterion}. This allows us to divide the surfing regime into two distinct behaviours: the `transient' surfing seen, for example, in the $\mathcal{T}=30$ case of figure \ref{fig:bob_v_surf}, where the initial velocity is sufficient to outpace the wave, but not to continue to do so for a long while, and `persistent' surfing like that of the $\mathcal{T}=40$ surfer, where the gel will enter a steady deswell-reswell cycle for some time, slowly losing fluid until it returns to a bobbing regime. A phase diagram illustrating the surfing and bobbing regimes is produced in figure \ref{fig:surfing_breakdown}.

    \section{Conclusions}
    We have introduced a new stroke mechanism for the design of artificial microswimmers that employs the interaction of chemically-responsive hydrogels with an oscillating background concentration field to drive non-reciprocal deswelling and reswelling. Even though the stimulus used to force the response of the hydrogel is periodic and entirely reversible in time, asymmetry in the responses of two spherical shells of hydrogel formed from gels with differing response characteristics results in a stroke action that is not reversible in time, permitting a net locomotion through a viscous fluid background. The advantages of such a design are manifold: though artificial swimmers have been constructed in the past that use spheres connected by elastic connections \citep{najafi-2004_simple_spheres}, or that operate by transporting fluid from one vessel to another \citep{avron-2005_pushmepullyou_swimmer}, this design requires no actuation machinery to be built in to the swimmer itself, endowing it with an intrinsic simplicity. All actuation occurs externally, and there is no need for an internal power source. Hydrogels are easy to fabricate and their properties can easily be tuned to specific applications, and are, by their very nature, flexible and soft.

    In order to model the response of these swimmers to an oscillating chemical field, we first developed a continuum-mechanical model that couples the dynamics of an oscillating reaction with a chemically-responsive gel and external fluid flow. Each aspect of this model exhibits complex feedback behaviour: for example, increasing chemical concentration leads to deswelling of the gel, which in turn can affect the rate of chemical reaction, or deswelling of the gel drives outwards fluid fluxes which can `wash away' chemical solutes and hence affect the rate of deswelling. In section \ref{sec:bz_gel_nd} we showed how the nature of these complicated interplays can be described by two non-dimensional parameters: a diffusion parameter $\mathcal{D}$ that measures the rate of solute diffusion relative to polymer scaffold reconfiguration and a Damk\"ohler number $Da$ representing the rate of reaction relative to gel deswelling or reswelling. In section \ref{sec:coreshell}, we introduce a gel geometry where we quantify the fluxes of fluid into and out of the gel in this framework, and we show how the swelling and deswelling fluxes for a gel undergoing oscillations in a periodic, pulsatile, chemical concentration field are not exact reverses of one another, and that self-oscillating gels provide a means by which a reciprocal input signal (in this case the reaction dynamics of the Brusselator model) can be converted into a non-reciprocal output (the morphology of the gels or fluid flows driven by swelling and drying).

    This inherent asymmetry was then used to design a two-sphere swimmer in section \ref{sec:purcell_swimmer}, where we have solved for the Stokes flow surrounding a pair of gels joined by a rigid connection with flows either into or out of the spheres that represent swelling or drying fluxes. We can use the modelling of the previous section to quantify these fluxes, and therefore deduce the forces on each sphere from viscous drag as the swimmer translates as a single unit through a background quiescent fluid and each sphere `pushes' or `pulls' the other by driving water in or out. Given the instantaneous flow field and translation velocity, we design a simple stroke based on the famous Purcell three-link swimmer where there is a net drifting motion in one preferred swimming direction that is second-order in the ratio of sphere radii to separation (assumed small). This proof-of-concept can then be converted into a more physically realisable swimmer by adding asymmetry between the two spheres, for example by tuning their response rates or degrees of deswelling. We also showed that analytic results can be found for the instantaneous and drift velocities by making the approximation that the sphere radii evolve like decaying exponentials, an assumption that is motivated by the form of the governing equation for their water and polymer content. This allows us to understand the influence of material properties and geometry on swimming speeds.

    Finally, we consider the interaction of responsive gel swimmers with spatially varying chemical concentration fields that take the form of repeated waves of high stimulus concentration and low stimulus concentration. Numerical simulations show two distinct behaviours are possible in this case -- either a simple `bobbing' motion that occurs when gels periodically swell and deswell, as analysed in section \ref{sec:in_phase}, and drift steadily as a result of intrinsic asymmetry, or a `surfing' motion when instantaneous velocities are higher and gels can outpace the motion of a chemical wave. In the former case, the direction of travel is a result of the geometry and gel properties of the swimmer itself, whilst in the latter case, swimmers travel with the periodic travelling wave, potentially at speeds much higher than the drift velocity. We present a simplified analytical model that explains the mechanism underlying surfing behaviour, and illustrating how gels are only able to surf for a limited period of time before exhausting their supply of water and resorting to a bobbing motion. In our models of these behaviours, however, the chemical kinetics and swimmer motion are entirely decoupled, with the background oscillating reaction waves not at all influenced by the release of water from swimmers as they deswell, nor do the flows generated affect the form of the chemical waves. This simplification, valid for especially small swimmers or very stable reaction-diffusion waves, could eventually be relaxed to drive more complex behaviours, where collective motion arises as the swimmers influence their local environment, a potentially rich avenue for future research built on the adaptable model of section \ref{sec:bzgel}.

    \begin{acknowledgments}
        This work was supported by the Leverhulme Trust Research Leadership Award `Shape-Transforming Active Microfluidics' (RL-2019-014) to TDMJ. JJW thanks Ellen Jolley and Danny Booth for helpful discussions that led to the spherical annulus geometry of section \ref{sec:coreshell}, and to two anonymous referees whose comments have significantly improved the study.
    \end{acknowledgments}

 \section*{Data availability}
    The code used to generate the figures in this article is openly available \citep{supp_data}.

    \appendix
    \section{Reaction-diffusion waves in the 1D Brusselator reaction}
    \label{app:bz_waves}
\begin{figure}
        \centering
        \includegraphics[width=0.9\linewidth]{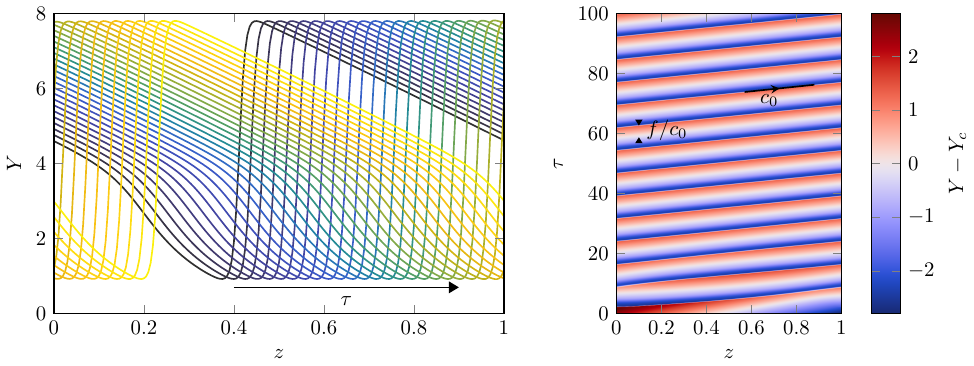}
        \caption{Plots of an example solution to the reaction-diffusion system of equation \eqref{eqn:bz_reactdiff} with $Da=1$ and $\mathcal{D} = 10^{-3}$ and an initial condition $X\equiv 0$, $Y=10(1-z)$ in a periodic box $0\le z \le 1$. On the left, plots of the travelling wave once it has settled into steady translation, and on the right a space-time diagram of the same results. In the right-hand plot, $Y_c=5$, and the different-coloured bands show how chemical concentration at any point oscillates between values above and below this critical value, with each band having a duration $f/c_0$ and propagating at speed $c_0$.}
        \label{fig:bz_waves}
    \end{figure}

    The BZ reaction is most known in the chemical literature for the intricate patterns that arise when chemical concentration fields are allowed to vary spatially and travelling waves of chemical concentration arise \citep{murase-2008_design_gel}. These waves can also be seen in reaction-diffusion models based on the Brusselator, and here we show how the chemical kinetics we use in the present study can give rise to wavelike solutions. In one spatial dimension $z$, the governing equations for the concentration of species $X$ and $Y$ in a solvent bath are
    \begin{subequations}
        \begin{align}
            \pdv{X}{\tau} &= \mathcal{D}\pdv[2]{X}{z} + Da\left[A + X^2 Y - (1+B)X\right], \\
            \pdv{Y}{\tau} &= \mathcal{D}\pdv[2]{Y}{z} + Da\left[B X - X^2 Y\right],
        \end{align}
        \label{eqn:bz_reactdiff}%
    \end{subequations}
    where $z$ is scaled with the poroelastic lengthscale $a_0$, $Da$ is the Damk\"ohler number relating poroelastic and reaction timescales, and $\mathcal{D}$ relates molecular diffusivity to poroelastic diffusivity, as seen above in equation \eqref{eqn:sys_react_diffuse}. Unlike earlier, the reaction is taking place in a fluid bath with no background imposed flow, and not in the gels themselves. This reaction-diffusion system can admit propagating wavelike solutions for certain values of the parameters $A$, $B$, $Da$ and $\mathcal{D}$ \citep{deller-1991_asymptotic_chaos}, as illustrated in figure \ref{fig:bz_waves}, and hence a chemically-responsive hydrogel at a fixed position $z$ in the background fluid will periodically deswell and reswell in response to the oscillating chemical concentration as waves pass it by. This behaviour becomes more complicated when the gel itself moves as a response to this deswelling, and therefore may interact in a more complex manner with the chemical waves.

    This one-dimensional model provides the simplest possible model for reaction-diffusion waves experienced by the small microswimmers, with the deswelling and reswelling of the gel being negligible in affecting the reaction dynamics. This decouples the chemical problem from the swimming problem, and is a reasonable assumption provided that the swimmers are small, and so fluid imbibition and expulsion do not destabilise the wavelike patterns in $X$ and $Y$ as they pass. Likewise, any effects in other spatial dimensions are ignored here on the assumption that the larger-scale structures (such as radially-spreading waves in 2D) appear locally planar in the frame of a swimmer.

    \section{Influence of non-constant permeability}
    \label{app:permeability}
\begin{figure}
        \centering
        \includegraphics[width=0.9\linewidth]{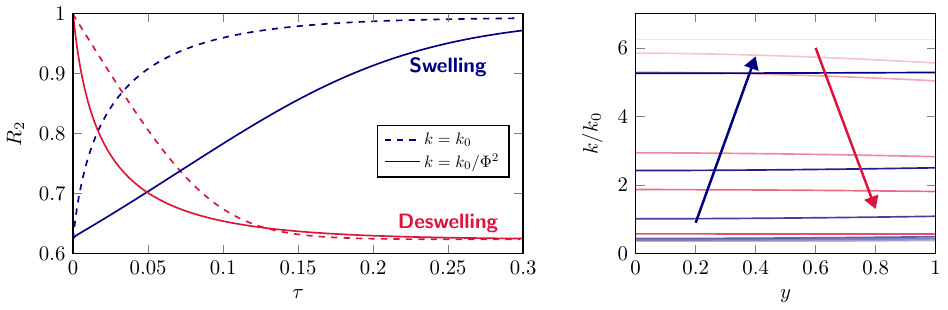}
        \caption{On the left, plots of the outer radius when swelling and drying using the same parameters as in figure \ref{fig:exp_fit} for both constant and varying permeability. Assuming non-constant permeability $k \propto \Phi^{-2}$, the permeability profile throughout the scaled thickness of the gel is plotted at different snapshots in time in the right-hand plot.}
        \label{fig:permeability_comparison}%
    \end{figure}

    Though the modelling in this study uses a constant permeability for analytic simplicity, it is apparent that the permeability of a hydrogel will depend on its porosity, with a greater pore space leading to a larger permeability. This is often modelled using the Kozeny-Carman relation, wherein
    \begin{equation}
        k = k_0 \frac{\varphi^3}{(1-\varphi)^2} = k_0 \frac{(1-\phi)^3}{\phi},
    \end{equation}
    where $\varphi$ is the porosity $1-\phi$. Since the hydrogels we are modelling in the present study have a high water content, $\phi_{00} \ll 1$, and so, at leading order, $k \propto \Phi^{-2}$, where $\Phi = \phi/\phi_{00}$. Figure \ref{fig:permeability_comparison} shows the effect of taking a non-constant permeability on the swelling and deswelling dynamics of a spherical annulus as considered in section \ref{sec:coreshell} versus the constant permeability used in this work. When deswelling, shrinkage initially happens faster due to the large pore spaces, before eventually the permeability drops as the gel becomes drier and dynamics are slower. Conversely, when swelling, growth of the outermost radius $R_2$ is initially slow, since the gel is dried, before the rate increases as the pore gaps open up. However, the same monotonic growth and shrinkage behaviour is seen irrespective of the permeability model chosen, and we therefore commit to using a constant permeability for illustrative purposes in the present work. 

\end{document}